\documentclass[]{aa}
\usepackage{graphicx}
\usepackage{tabu}
\usepackage{longtable,ltcaption}
\usepackage{tabularx,ragged2e,booktabs}
\usepackage[normalem]{ulem} 

\usepackage{txfonts}

\usepackage{natbib}
\usepackage[colorlinks=true, linkcolor=blue, citecolor=blue, filecolor=blue, urlcolor=blue]{hyperref}
\usepackage{placeins}

\usepackage[dvipsnames]{xcolor}

\titlerunning{Radio-Continuum Spectra of Pulsars}
\authorrunning{Abadi et al.}

\begin{document}

\title{Radio-Continuum Spectra of Pulsars with Free-Free Thermal Absorption}

 \author{Mario G. Abadi \inst{1,2}\href{https://orcid.org/0000-0003-3055-6678}{\includegraphics[scale=0.8]{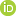}}\thanks{E-mail:
mario.abadi@unc.edu.ar}
\and Gabriela Castelletti\inst{3}\href{https://orcid.org/0009-0002-0134-2064}{\includegraphics[scale=0.8]{LEGACY/Figures/orcid.png}}
\and Namir E. Kassim\inst{4}\href{https://orcid.org/0000-0001-8035-4906}{\includegraphics[scale=0.8]{LEGACY/Figures/orcid.png}}
}

\institute{CONICET-Universidad Nacional de C\'{o}rdoba, Instituto de
Astronom\'{i}a Te\'{o}rica y Experimental (IATE),
Laprida 854, X5000BGR,C\'{o}rdoba, Argentina
\and
Observatorio Astron\'{o}mico, Universidad Nacional de C\'{o}rdoba,
Laprida 854, X5000BGR, C\'{o}rdoba, Argentina
\and 
Instituto de Astronom\'ia y F\'isica del Espacio (IAFE), Ciudad Universitaria - Pabellón 2, Intendente Güiraldes 2160 (C1428EGA), Ciudad Aut\'onoma de Buenos Aires - Argentina
\and
Remote Sensing Division, Naval Research Laboratory, Code 7213, 4555 Overlook Ave SW, Washington, DC, 20375, USA
}

\date{Received May XX, 2023; accepted Month XX, 2023}

\abstract
{The radio continuum spectra of pulsars (PSRs) exhibit 
a wide variety of shapes, 
 that are interpreted as 
pure and broken power laws, power laws with turnovers or cut-offs, and logarithmic-parabolic profiles. 
A notable fraction of these have well-defined power laws with $\nu^{-2.1}$ exponential turnovers, indicative of free-free thermal absorption along the line-of-site. We analyse a sample of 63 PSRs with such spectral shapes, compiled from four previously published studies, to  investigate their    statistical properties. We normalise each spectrum to a characteristic frequency and flux density of its own, facilitating a consistent treatment across the four sub-samples. 
We show these two fitted parameters are correlated by a power law, with its slope  reflecting the median spectral index ($\alpha\sim -2.0$) of PSR  emission.
We found that the turnover frequencies in our sample are typically high, clustering around 558~MHz, implying notably high emission measures (EM$\sim 10^{5}$~pc~cm$^{-6}$) for an inferred thermal absorbing medium with electron temperature of $T_{\mathrm{e}}=8000$~K.
Moreover, by combining these EM with dispersion measures (DM) derived from pulse time delays, we break the degeneracy between electron density and path length of the absorbers. This reveals a discrete near-in population of absorbers characterised by small sizes (\(L \sim 0.1\,\text{pc}\)) and high electron densities (\(n_{\mathrm{e}} \sim 10^{3}\,\text{cm}^{-3}\)), which exhibit a clear size-density anticorrelation  reminiscent of that observed in Galactic and extragalactic H\,{\sc ii} regions.}

\keywords{ISM: general; ISM: structure; Radio continuum: ISM; (Stars:) pulsars: individual}
\maketitle

\section{Introduction}
\label{sec:intro}

The study of pulsar (PSR) radio continuum spectra began in the 1970s, shortly after their discovery. Early works compiled flux density measurements across a range of frequencies, laying the groundwork for the first characterisations of their spectral behaviour. A seminal contribution came from 
\citet{sieber1973}, who analysed spectra of PSRs using the observational data then available. Although each source was typically sampled at only a few frequencies, the study revealed that many PSR spectra followed a  power-law trend, hinting at a  relatively simple underlying emission mechanism. 
  \citet{sieber1973} also noted in several cases that the spectra showed low-frequency turnovers, which he attributed to either synchrotron self-absorption (SSA) or thermal free-free absorption (FFA). 
These early findings highlighted the diversity of PSR spectral behaviour. 

As observational capabilities expanded--both in terms of frequency coverage and sensitivity--it became increasingly evident that deviations from a simple power-law model are more common than initially thought. Nowadays, a variety of  spectral features are observed, 
such as broken power laws,  flattening, and turnovers, spanning a broad frequency range from
$\sim$30~MHz up to and  $\sim$1~GHz \citep[e.g.,][]{swainston2022,lee2022}. These advances have been driven by dedicated pulsar surveys accomplished with  new-generation radio facilities, ranging from early initiatives such as the Parkes Southern Pulsar Survey \citep{manchester1996} to more recent ones as  
The GMRT High Resolution Southern Sky Survey for Pulsars and Transients \citep[GHRSS,][]{bhattacharyya2016}, 
the LOFAR Tied-Array All-Sky Survey \citep[LOTAAS,][]{sanidas2019}, 
The Southern-sky MWA Rapid Two-metre pulsar survey \citep[SMART,][]{bhat2023}, 
The Green Bank 820~MHz Pulsar Survey \citep{mcewen2024},
The FAST Galactic Plane Snapshot Survey \citep[GPPS,][]{han2025}, and 
The NenuFAR Pulsar Blind Survey \citep[NPBS,][]{brionne2025}, among others. 
Despite this growing body of data, the physical origin of the observed spectral deviations remains under debate--whether they stem from intrinsic properties of the pulsar emission mechanism or arise from extrinsic propagation effects through intervening media \citep[e.g.,][]{jankowski2018}. 

Among the various spectral features, exponential low-frequency attenuation consistent with FFA has gained increasing support as a tracer of ionised gas along the line of sight. 
Although FFA had already been proposed by \citet{sieber1973} to account for MHz turnovers, more recent studies 
\citep[e.g.,][]{kijak2011-binary,lewandowski2015,rajwade2016,kijak2017,basu2018,kijak2021} have extended this interpretation to include higher-frequency turnovers, even reaching the GHz regime. 
These absorption signatures are thought to originate in diverse environments, such as  filaments in supernova remnants (SNRs), bow shocks from pulsar wind nebulae (PWNe), H\,{\sc ii} regions, or dense stellar winds in binary systems. In this context, modelling PSR spectra with exponential absorption functions provides a potentially powerful tool for probing the physical properties of the intervening ionised medium. In particular, the turnover frequency is sensitive to a combination of electron density, path length, and electron temperature via the emission measure (EM). However, different combinations of these parameters can result in similar turnover frequencies, making it difficult to uniquely constrain the physical conditions of the absorber. One promising strategy to mitigate this degeneracy involves complementing spectral fits with independent constraints from dispersion measures (DM)--induced pulse time delays, which trace the total column density of free electrons along the line of sight (e.g., \citealt{rajwade2016,kijak2021,rozko2021}). Furthermore, an additional  complication arises from the use of different mathematical forms and conventions in the fitting equation of the spectra across the literature, which hinders direct comparisons between studies.  Adopting a unique fitting equation  facilitates  comparisons across multiple samples, ultimately leading to insightful interpretations of the structures responsible for PSR absorption features. 

This paper presents an analysis of PSR radio continuum spectra  previously interpreted as bearing signatures of FFA. In Sect.~\ref{sec:data-analysis}, we employ the spectral modelling approach developed by \citet{abadi2024}--originally designed to homogenise SNR spectra affected by thermal absorption--to our sample of PSRs with turnovers extending up to GHz frequencies. In Sect.~\ref{sec:correlations}, we examine correlations and differences in the fitted spectral parameters for PSRs and SNRs. The  implications for the physical nature of the absorbing environments  are discussed in Sect.~\ref{sec:absorbing-medium}. Finally, Sect.~\ref{sec:conclusions} provides a summary of our results. 

\section{Parameterisation and similarity in radio-continuum PSR spectra}
\label{sec:data-analysis}

The database used in our study comprises radio continuum spectra of 63 Galactic pulsars that span a broad frequency range from 20~MHz to 10~GHz.
Among them, 15 were taken from \citet{kijak2017} (hereafter K17), 25 from 
\citet{jankowski2018} (J18), 10 from \citet{kijak2021} (K21),  and 13 from \citet{swainston2022} (S22).\footnote{We notice that \citet{swainston2022} refers to an open-source repository that centralises pulsar flux density measurements. The data included in our paper correspond to the catalogue update available at the time of writing.} 
These spectra are systematically modelled by a power-law related to the PSR's intrinsic radio emission with an exponential term that accounts for thermal free-free absorption along the light of sight:

\begin{equation}
\centering
  S(\nu) = S_{0}\, \Big(\frac{\nu}{\nu_{0}}\Big)^{\alpha} \, \mathrm{exp}\Big[-\tau_{0}\,\Big(\frac{\nu}{\nu_{\mathrm{0}}}\Big)^{-2.1}\Big]. 
    \label{equation:flux}
\end{equation}

\noindent
Here, $S_{0}$ is a reference flux-density corresponding to an arbitrarily chosen reference frequency $\nu_{0}$, $\alpha$ is the power-law spectral index, and $\tau_0$ is the optical depth at $\nu_0$. The optical depth $\tau_0$ at frequency $\nu_{0}$, can be expressed as:

\begin{equation}
\tau_{0} = 8.235 \times 10^{-2} \, \Big(\frac{T_{\mathrm{e}}}{\mathrm{K}}\Big)^{-1.35} \, \Big(\frac{\nu_{0}}{\mathrm{GHz}}\Big)^{-2.1} \, \Big(\frac{\mathrm{EM}}{\mathrm{pc~cm^{-6}}}\Big) \, a(\nu_{0}, \mathrm{T}_{\mathrm{e}}),
\label{equation:tau-modified}    
\end{equation}

\noindent
where $T_{\mathrm{e}}$ is the mean electron temperature of the absorbing medium, $a(\nu_{0}, T_{\mathrm{e}})$ is the Gaunt factor correction--typically assumed to be of order unity \citep{wilson09}-- and EM is the emission measure, defined as

\begin{equation}
    \mathrm{EM}=\int_{0}^{L} n_{\mathrm{e}}^{2}\, dx,
    \label{equation:EM}
\end{equation}

\noindent
with $n_{\mathrm{e}}$ the free electron density and $L$ the path length through the absorbing medium.

Different authors have adopted slightly varied forms for Eq.~\ref{equation:flux}, either adjusting the reference frequency or by organising parameters differently. However, for PSR spectra affected by thermal free-free absorption in the intervening medium, Eq.~\ref{equation:flux} consistently remains a three-parameter ($S_{0}$, $\alpha$, $\tau_{0}$) model for fitting measured flux densities, whereas the reference frequency $\nu_{0}$ is arbitrarily fixed by each author. Even though transitioning between parameter values in different expressions is straightforward, discrepancies in parameterisation complicate comparisons. 
Following  \citet{abadi2024}, we have applied their parameterisation for SNR continuum spectra to PSRs, i.e., we rewrote Eq.~\ref{equation:flux} 
in terms of a characteristic frequency, $\nu_{\ast}$, and flux density, $S_{\ast}$: 

\begin{equation}
\centering
S(\nu)=S_{\ast} \Big( \frac{\nu}{\nu_{\ast}} \Big) ^{\alpha} \mathrm{exp} \Big[ -\Big(\frac{\nu}{\nu_{\ast}}\Big)^{-2.1} \Big].
    \label{equation:sast}
\end{equation}

The characteristic frequency $\nu_{\ast}$ is related to the radio spectral index $\alpha$ and the turnover frequency  $\nu_{\mathrm{to}}$  (i.e., the frequency at which $S(\nu)$ reaches its maximum value from the fitting process) through the expression $\nu_\mathrm{to}/\nu_{\ast}=(-2.1/\alpha)^{1/2.1}$, while $S_{\ast}$ represents the spectral normalisation. This alternative formulation offers the key advantage of being independent of any reference frequency $\nu_{0}$.  Indeed, by comparing Eq.~\ref{equation:flux} and Eq.~\ref{equation:sast}, it follows that  $\nu_{\ast}=\tau_{0}^{1/2.1}\, \nu_{0}$ and $S_{\ast}=S_{0}\,(\nu_{\ast}/\nu_{0})^{\alpha}$,  implying that the characteristic parameters 
$\nu_{\ast}$ and $S_{\ast}$ remain invariant under changes of reference frequency. Using these expressions along with the fitting parameters reported by K17, J18, K21, and S22, we derived the values of  $S_{\ast}$ and $\nu_{\ast}$ listed in Table~\ref{tab:long}.  
Given the relationship between the characteristic frequency and the free-free optical depth, Eq.~\ref{equation:tau-modified} can be rewritten as: 

\begin{figure}

 \includegraphics[width=0.45\textwidth,height=0.19\textwidth]{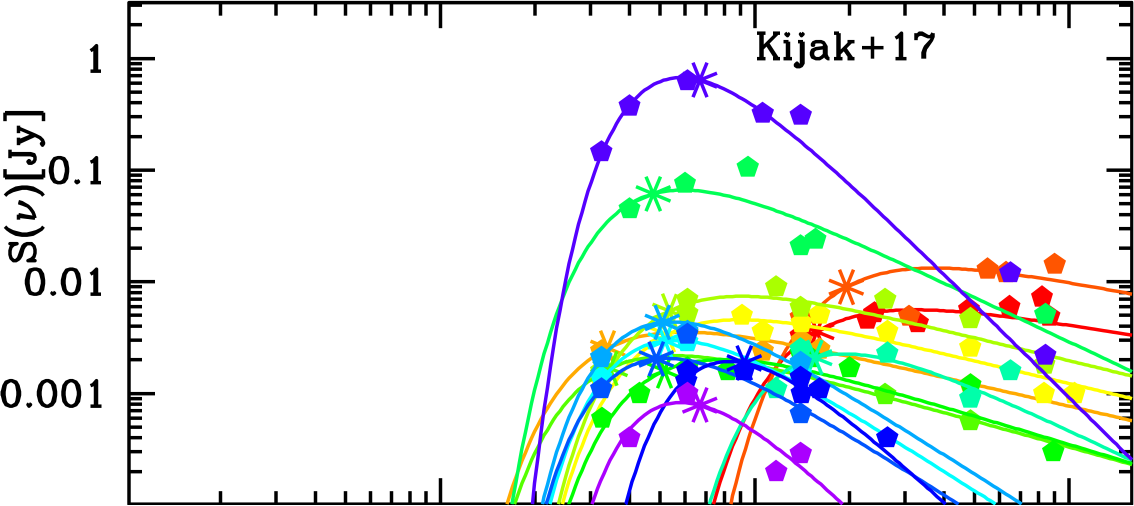}
   \includegraphics[width=0.45\textwidth,height=0.19\textwidth]{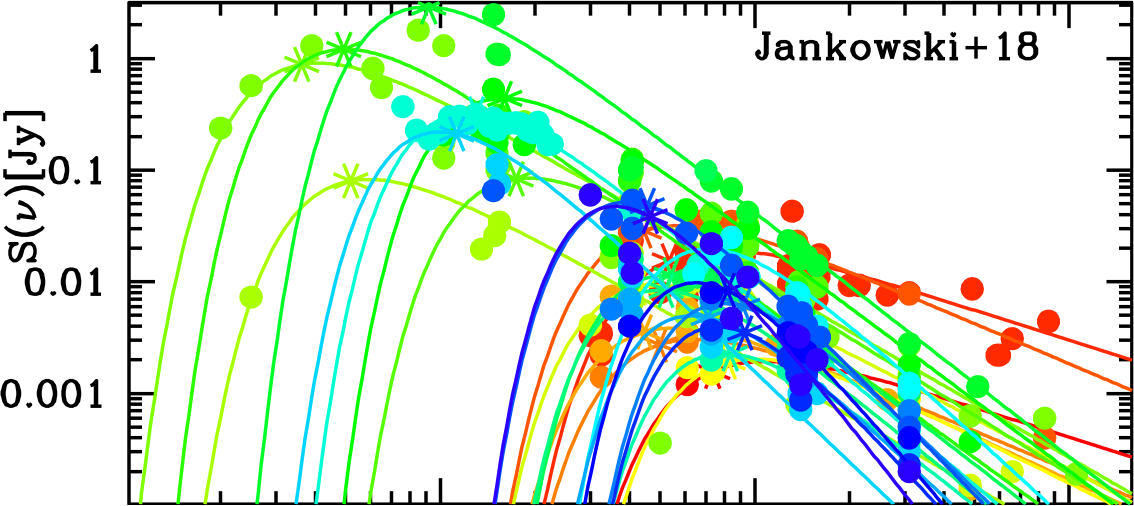}
   \includegraphics[width=0.45\textwidth,height=0.19\textwidth]{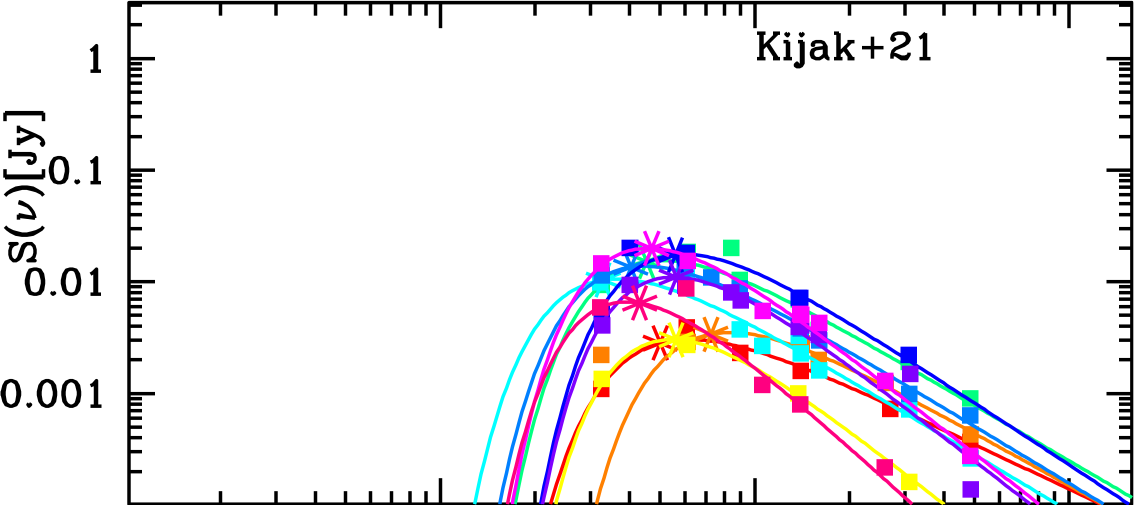}
\includegraphics[width=0.45\textwidth,height=0.23\textwidth]{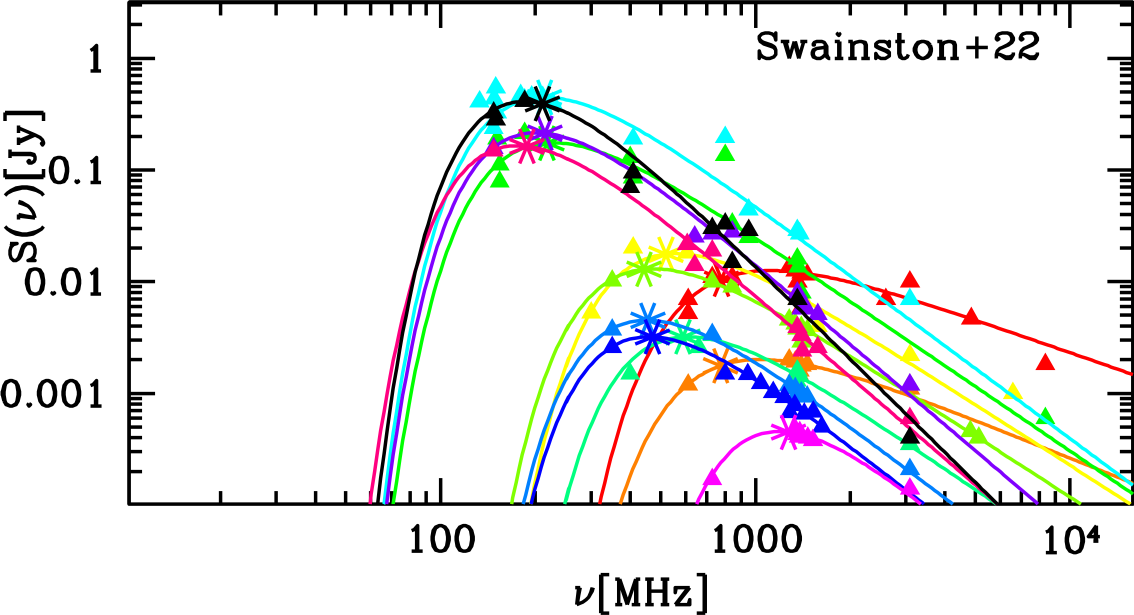}
  \caption{Radio flux density measurements (coloured filled symbols) vs. frequency for the four PSR subsamples analysed in our study. The panels, from top to bottom, display 15 spectra from K17, 25 from J18, 10 from K21, and 13 from S22. The solid coloured lines represent the best-fit models derived from  Eq.~\ref{equation:sast}, using  spectral indices reported in the literature referenced in our work (see Table~\ref{tab:long} for the parameter values). Asterisk symbols mark the characteristic frequency, $\nu_{\ast}$, for each individual spectrum.
}
\label{figure:fig1}
\end{figure}

\begin{equation}
\nu_{\ast} = 0.305 \, \mathrm{GHz} \, \Big(\frac{\mathrm{EM}}{\mathrm{pc~cm^{-6}}}\Big)^{1/2.1} \, \Big(\frac{T_{\mathrm{e}}}{\mathrm{K}}\Big)^{-0.64}. 
   \label{equation:nuast}
\end{equation}
This expression shows that, assuming a fixed electron temperature, our fitting parameter $\nu_{\ast}$ provides a direct estimate of EM (see Sect.~\ref{sec:absorbing-medium}).

\begin{figure}
\includegraphics[width=0.45\textwidth,height=0.19\textwidth]{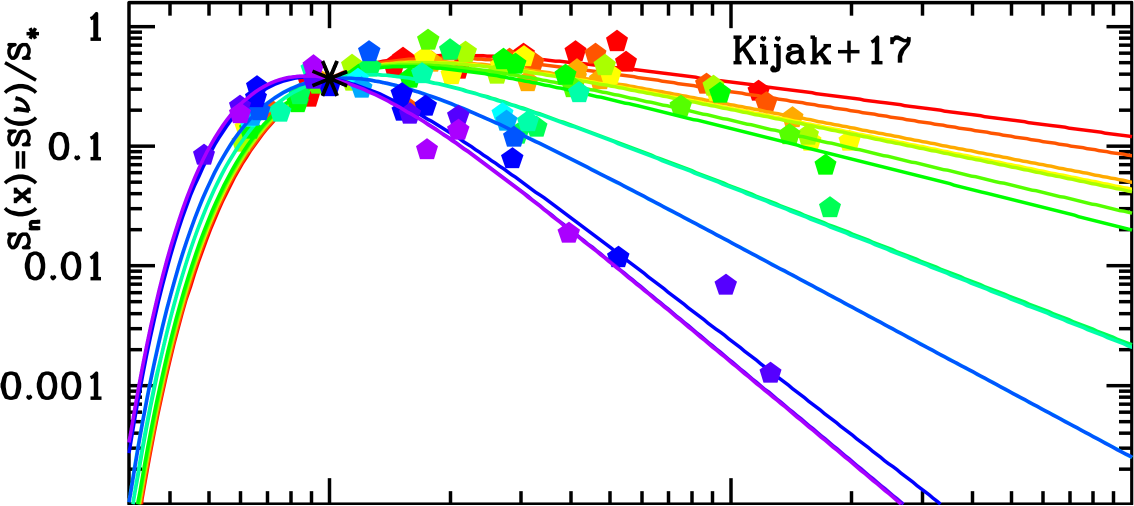}
\includegraphics[width=0.45\textwidth,height=0.19\textwidth]{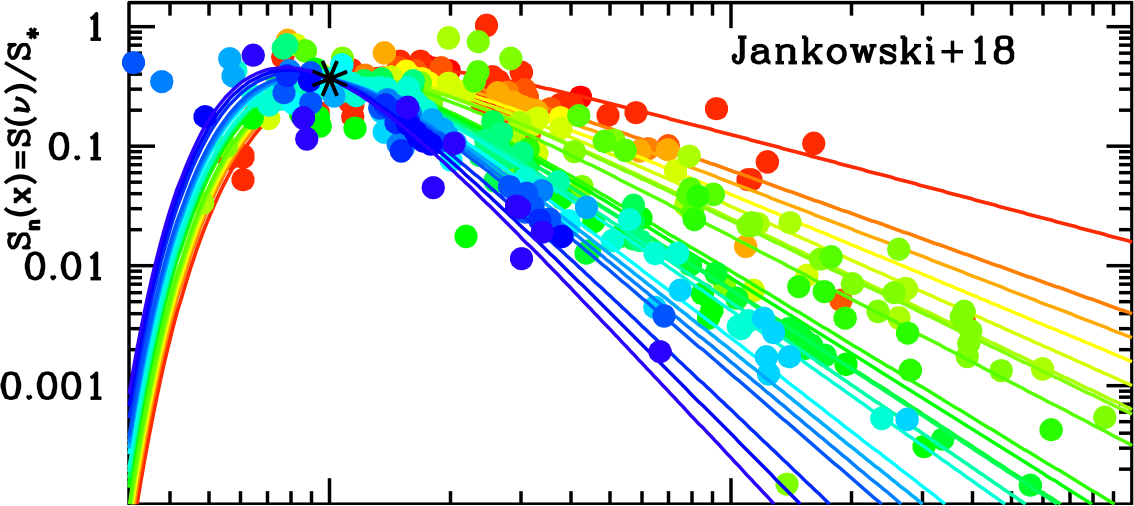}
\includegraphics[width=0.45\textwidth,height=0.19\textwidth]{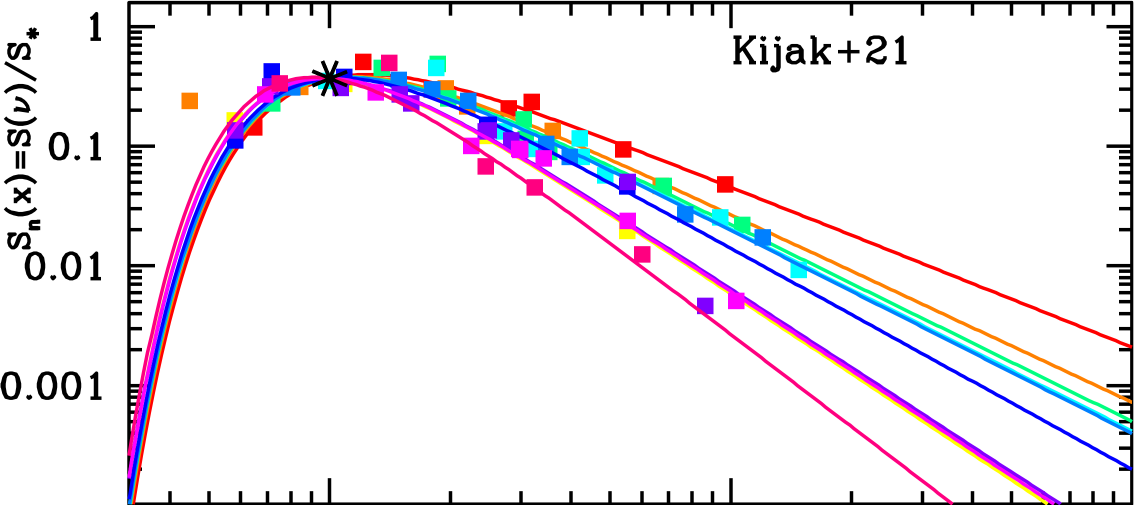}
   \includegraphics[width=0.45\textwidth,height=0.23\textwidth]{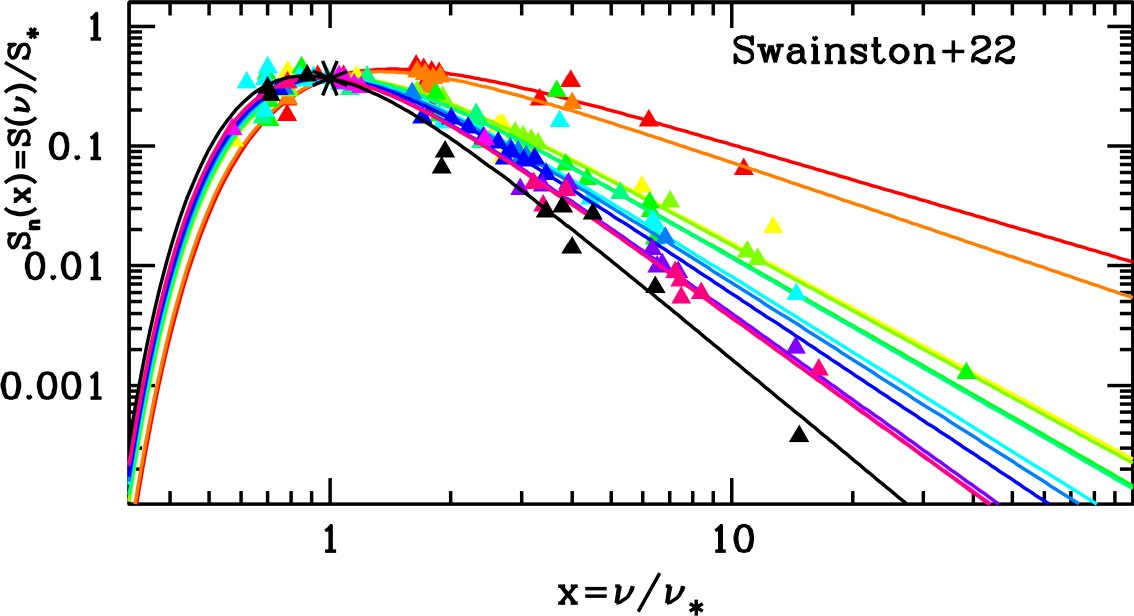}
  \caption{Normalised radio flux density  as a function of normalised frequency for our PSR sample, comprising (from top to bottom) 15 objects reported in K17, 25 in J18, 10 in K21, plus 13 in S22. 
  Each PSR within the subsamples is colour-coded individually. The solid lines represent the fitted models, with parameters listed in Table~\ref{tab:long}, and the black asterisk indicates the characteristic frequency, $\nu_{\ast}$.
This normalisation process highlights the similar exponential drop-off in the spectra associated with $\nu_{\ast}$ values, arranging PSR spectra in increasing order based on the slope of their emission. 
}
\label{figure:fig2}
\end{figure}

The panels in Fig.~\ref{figure:fig1}   separately display the radio continuum spectra of each pulsar subsample (K17, J18, K21, and S22) considered in our analysis. To ensure that only spectra with robust fits were included, we selected those for which the chi-square statistic, 
$\chi^{2}={\Sigma\,(\mathrm{log}S_{\mathrm{obs}}-\mathrm{log}S_{\mathrm{fit}})^{2}/N}$, is lower than $\sim 0.06$, where 
$S_{\mathrm{obs}}$ and $S_{\mathrm{fit}}$ are the observed and modelled flux densities, respectively, and $N$ is the number of data points used in the fit.
Importantly, rather than performing new fits to the compiled data, we adopted the fitting parameters already published by the authors of each subsample to derive $S_{\ast}$ and $\nu_{\ast}$ for Eq.~\ref{equation:sast}. This selection based on $\chi^{2}$ led to the exclusion of 21 poorly fitted spectra from S22, while all spectra from  K17 and K21 were retained. The J18 subsample, however, could not be subjected to this selection, as the individual flux density measurements were not published. In addition, as J18 provide only the turnover frequency and the spectral index, but not the fitted flux density scaling parameter (which sets the vertical normalisation of the spectrum), we recalculated the missing parameter using the spectral measurements compiled in the S22 repository. As both J18 and S22 compiled flux densities from the literature, we do not expect this procedure to introduce any systematic bias in the J18 spectra.
Also, it is acknowledged that, in the case of a few pulsars for which only a single measurement causes the spectrum to deviate noticeably from a power-law, these sources were tentatively included based on the available data. However, a definitive confirmation that their spectra indeed exhibit the characteristic signatures of FFA would require further investigation, which lies beyond the scope of this paper. \rm

\begin{figure}
   \includegraphics[width=0.45\textwidth,height=0.19\textwidth]{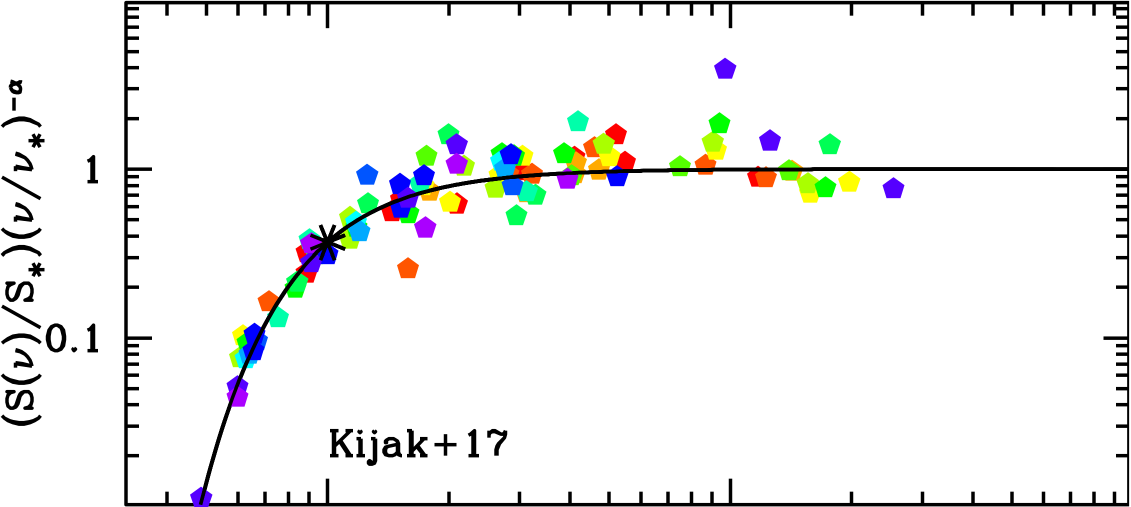}
   \includegraphics[width=0.45\textwidth,height=0.19\textwidth]{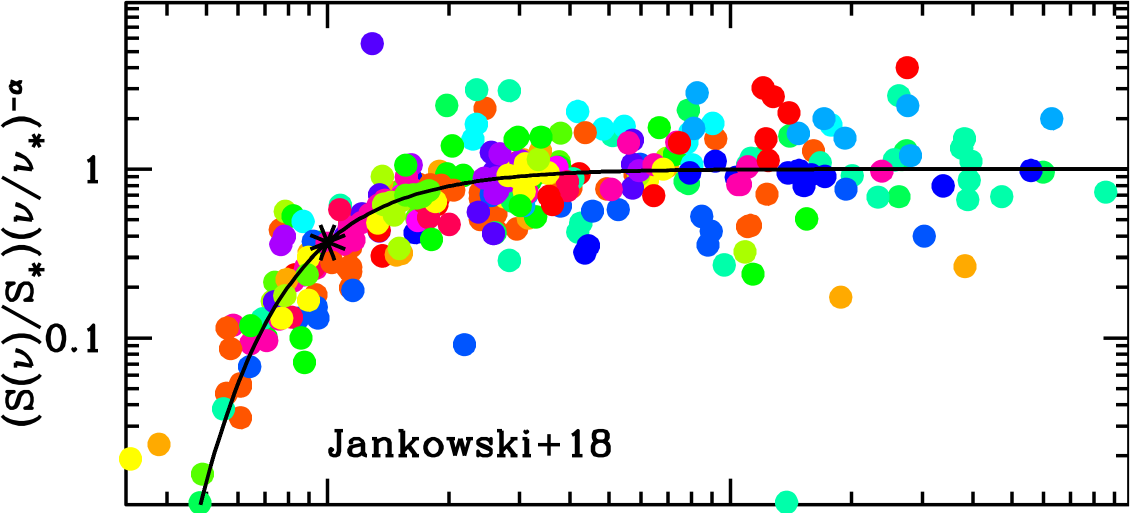}
   \includegraphics[width=0.45\textwidth,height=0.19\textwidth]{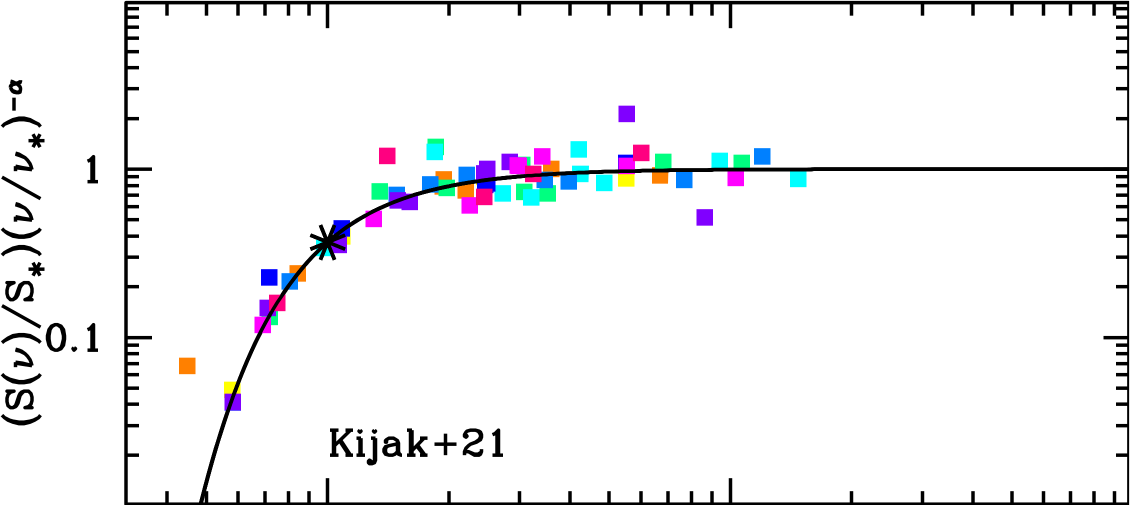}    \includegraphics[width=0.45\textwidth,height=0.23\textwidth]{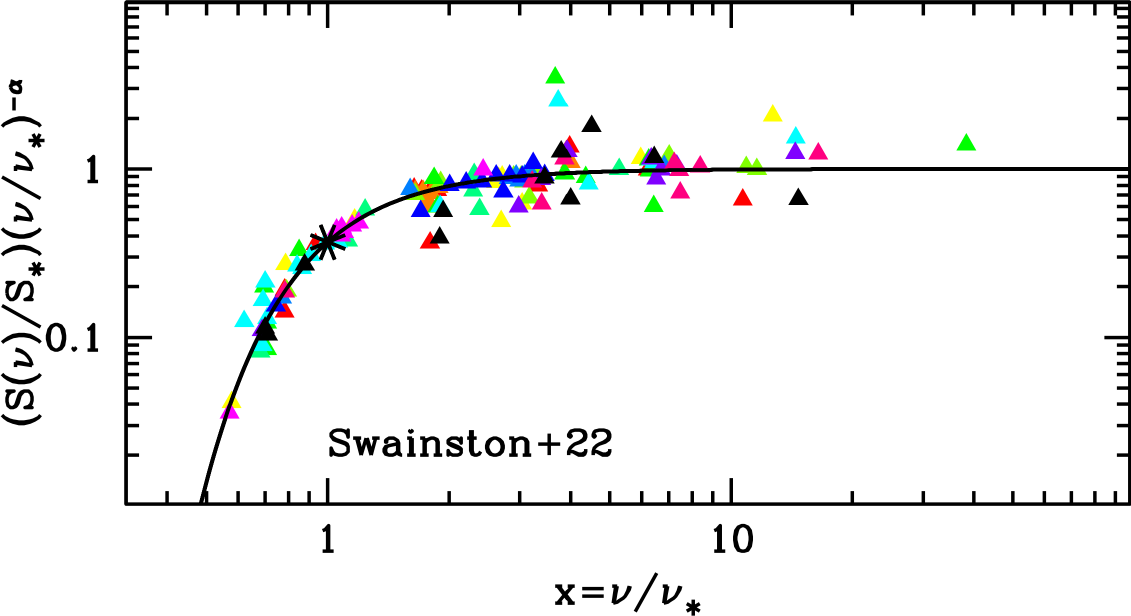}
 \caption{Radio flux density,   normalised by the characteristic flux and power-law emission,  plotted against  normalised frequency for the four subsamples in our study (see labels).  Each spectrum was fitted by the respective authors with an exponential drop-off with an exponent $-2.1$.  The curved solid line represents the normalised absorption function  $A(x)=\mathrm{exp}(-x^{-2.1})$
 and the asterisk symbol indicates the characteristic frequency, $\nu_{\ast}$.}
\label{figure:fig3}
\end{figure}

\begin{figure}
     \includegraphics[width=0.45\textwidth]{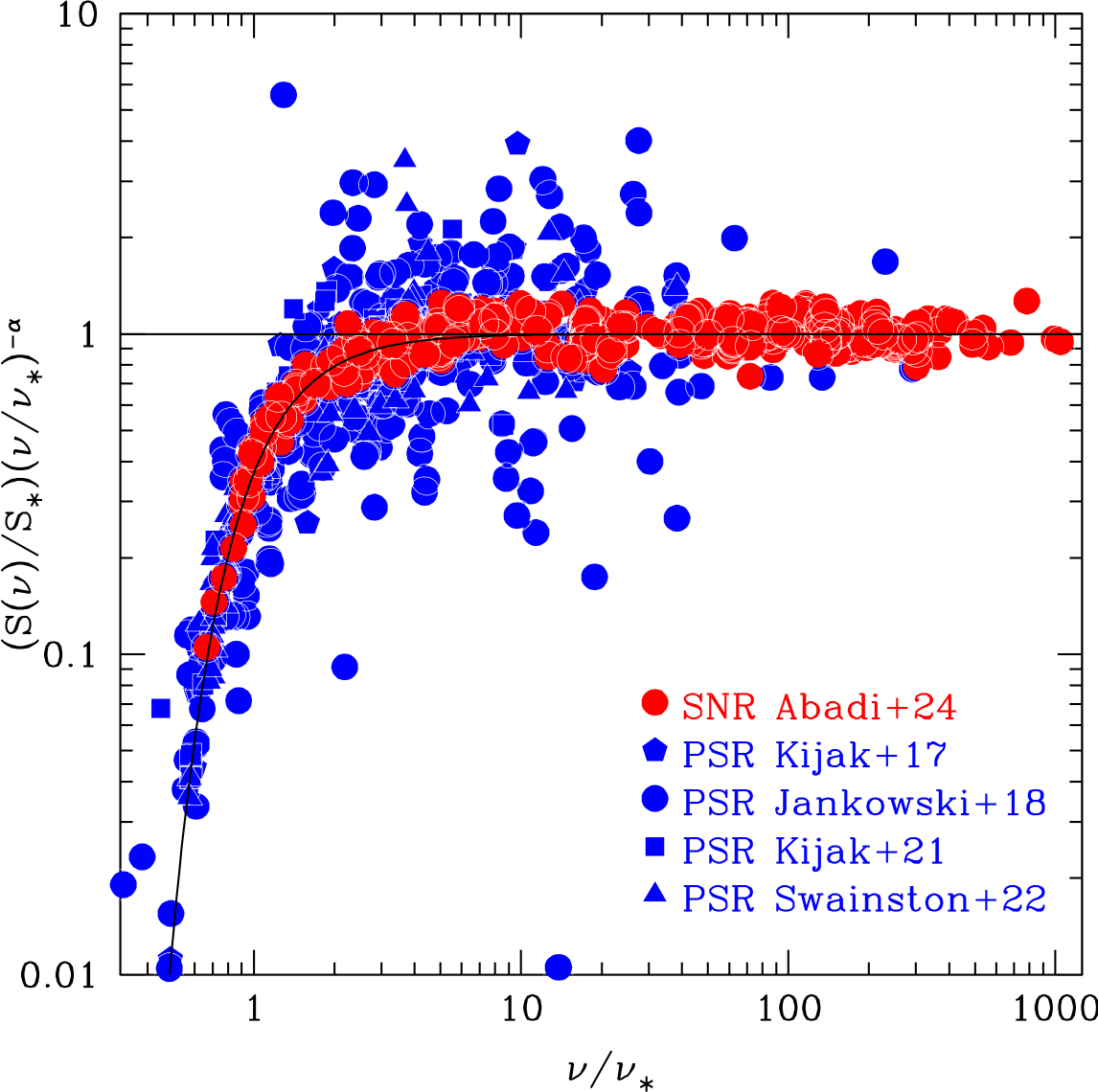}
    \caption{Same PSR data as plotted in Fig.~\ref{figure:fig3}, with the four subsamples merged (blue symbols). 
 Additionally, data points for the SNRs presented in \citet{abadi2024} are reproduced (red symbols). The graph highlights 
 the similarity not only between the PSR spectra but also between the PSR and SNR spectral behaviours.
The curved solid line is the normalised absorption $A(x)=\mathrm{exp}(-x^{-2.1})$.
}
    \label{figure:fig4}
\end{figure}

In Fig.~\ref{figure:fig2}, we illustrate our methodology by plotting the normalised flux density 
$S_{\mathrm{n}}(x)=S(\nu)/S_{\ast}=x^{\alpha}\, \mathrm{exp}(-x^{-2.1})$ as a function of the normalised frequency $x=\nu/\nu_{\ast}$ for all the
spectra depicted in Fig.~\ref{figure:fig1}. 
This new approach reveals a striking similarity in how pulsar spectra exhibit turnovers at frequencies $\nu < \nu_{\ast}$, despite  emitting  with varying radio spectral indice. This behaviour is further illustrated in Fig.~\ref{figure:fig3}, where all PSR radio spectra are fitted using a unified  absorption function $A(x)= \mathrm{exp}(-x^{-2.1})$, shown as the solid black curve. 
In \citet{abadi2024}, we demonstrated that SNR spectra often, but not uniformly, exhibit a drop-off at frequencies below 100~MHz 
due to thermal free-free absorption. Here, we emphasise that a similar feature also 
characterises the spectral turnovers of PSRs, occurring at around 558~MHz. We highlight this  property in Fig.~\ref{figure:fig4}, where we  plot the ratio between the normalised flux densities and the power-law emission component of each PSR spectrum (blue symbols) as a function of the normalised frequency. For comparison, we include the 12 SNR spectra from \citet{abadi2024} (red symbols). While both PSRs and SNRs follow the same exponential decay function $A(x)$, the scatter around the curve is noticeably larger for the 63 PSRs than for the SNRs. This discrepancy can be attributed to the more stringent selection criteria adopted by \citet{Castelletti2021} when constructing individual SNR spectra, rather than to any intrinsic difference in the spectral scatter between SNRs and PSRs. Moreover, the normalised flux measured at the lowest normalised frequency, $x \sim 0.5$, reaches a value of $\sim 0.01$ for PSRs, compared to $ \sim 0.1$ for SNRs. This behaviour arises directly from the relation 
$\nu_\mathrm{to}/\nu_{\ast}=(-2.1/\alpha)^{1/2.1}$, which leads to 
$\nu_{\ast} \sim \nu_{\mathrm{to}}$
for PSRs and  $\nu_{\ast} \approx \nu_{\mathrm{to}}/2$
for SNRs,  given  their different characteristic  power-law indices, $\alpha \sim -2.0$ and $-0.5$, respectively.

\section{Spectral parameters correlations }
\label{sec:correlations}
\begin{figure}
     \includegraphics[width=0.45\textwidth]{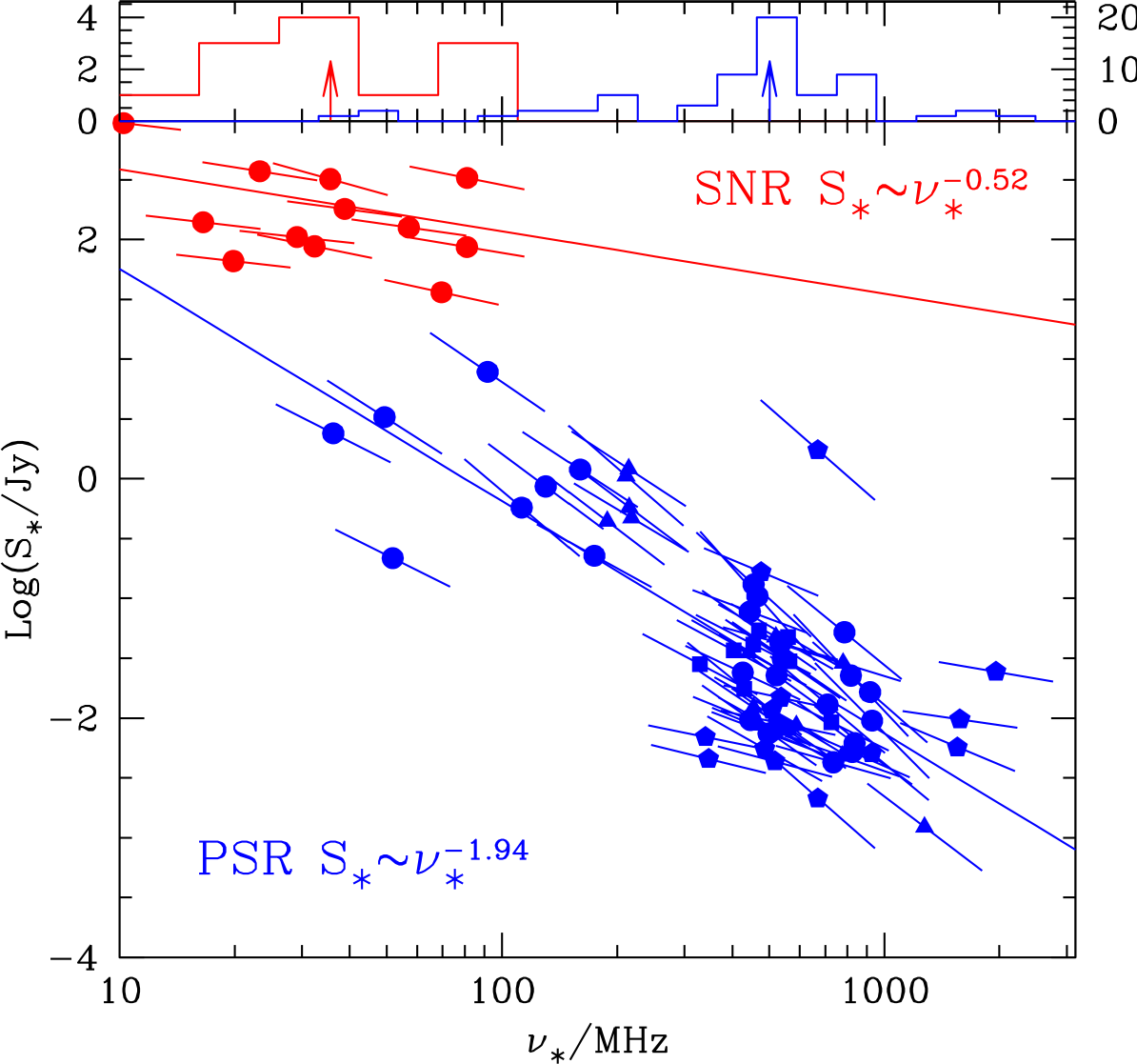}
    \caption{Relation between the characteristic flux, $S_{\ast}$, and frequency, $\nu_{\ast}$, parameters obtained from fitting the parameterised free-free thermal absorption model (Eq.~\ref{equation:sast}) for our sample of 63 PSRs. 
    For comparison, parameters calculated from the same model fit applied to 12 SNRs \citep{abadi2024} are also included. 
    All data points are shown with the same symbols and colours as in Fig.~\ref{figure:fig4}. 
    The large solid lines represent the best-fit linear relations  for the PSR (blue) and SNR (red) parameters, with logarithmic slopes of $-1.94$ and $-0.52$, respectively. 
    Short lines on each data point indicate the radio spectral index for each object, taken from K17, J18, K21, and S22 for PSRs and from \citet{abadi2024} for SNRs. The median values of these spectral indices are consistent with the slopes of the power-law correlation between $S_{\ast}$ and  $\nu_{\ast}$.  
    The distribution of $\nu_{\ast}$ values for PSRs and SNRs, with median values indicated by arrows, is shown in the blue and red histograms in the top panel. 
    }
    \label{figure:fig5}
\end{figure}

The proposed parameterisation of PSR spectra in terms of $S_{\ast}$ and $\nu_{\ast}$ (Eq.~\ref{equation:sast}) enables a  homogeneous treatment of the four subsamples (K17, J18, K21, and S22), allowing us to analyse collectively the 63 spectra included in our compiled sample. This unified approach provides a consistent framework to investigate the statistical properties of the full dataset and to explore potential correlations among the fitted quantities. In Fig.~\ref{figure:fig5}, we display the mutual relation between $S_{\ast}$ and $\nu_{\ast}$ for the PSRs  (blue symbols). For comparison, we also include (red symbols) the corresponding values for a set of 12 SNR radio spectra with low-frequency turnovers presented in \citet{abadi2024}. The top panel shows histograms of the $\nu_{\ast}$ distributions for both samples.
The PSRs exhibit a median $\tilde{\nu}_{\ast}=500$~MHz, which corresponds to a median $\tilde{\nu}_{\mathrm{to}}=558$~MHz. 
In contrast, the SNRs  show a significantly lower median $\tilde{\nu}_{\ast}=36$~MHz, equivalent to a turnover frequency of $\tilde{\nu}_{\mathrm{to}}=68$~MHz. 
The slope of the observed correlation between $\nu_{\ast}$ and $S_{\ast}$ primarily reflects the median of the  distribution of spectral indices. In particular, for our PSR sample  $\alpha$ values 
range from approximately  $-4.0 \lesssim \alpha \lesssim -0.5$ (see blue histogram in the top panel of Fig.~\ref{figure:fig6}) with a median of $ \tilde{\alpha}=-1.85$ in good agreement with 
the slope $\alpha=-1.94$ obtained from the best-fit models applied to the data points.  A similar trend is observed for the  SNR sample, where the spectral indices span the range $-1.0 \lesssim \alpha \lesssim -0.3$ (see red histogram in the top panel of Fig.~\ref{figure:fig6}),  with a median value of $\tilde{\alpha}=-0.46$, closely matching the best-fit power-law slope $\alpha=-0.52$. The $S_{\ast}$ values indicate the projected power-law flux density at $\nu_{\ast}$ in the absence of absorption, while the short line segments indicate its slope. 
A visual inspection shows that, qualitatively, the slopes of these segments follow approximately the correlation exhibited by the data points themselves.

We also explored possible correlations between the spectral index $\alpha$ and both $\nu_{\ast}$  and $S_{\ast}$, but found no significant trends, as shown in Figs.~\ref{figure:fig6} and \ref{figure:fig7}, respectively. In both figures, we maintain the same colour scheme as in Fig.~\ref{figure:fig5}, using blue for PSRs and red for SNRs. 
The intrinsic  emission from the source is described by the power-law function term in Eq.~\ref{equation:sast}.  Consequently, the luminosity  can be readily estimated as $\mathcal{L}=4\pi\,d^{2} \int_{\nu_{\mathrm{min}}}^{\nu_{\mathrm{max}}} S_{\ast} \, (\nu/\nu_{\ast})^{\alpha} \, d\nu$, where  $d$ represents the heliocentric distance to the source. We then extended our search for correlations between $\alpha$ and $\mathcal{L}$. 
One additional advantage of the parameterised spectral model (Eq.~\ref{equation:sast}) is the ability to
integrate flux density over fixed values of normalised frequency $x=\nu/\nu_{\ast}$, 
without the need to define $\nu_{\mathrm{min}}$ and $\nu_{\mathrm{max}}$ separately for each source. 
We used this feature to obtain an expression for the radio luminosity:  

\begin{figure}
  \includegraphics[width=0.45\textwidth]
{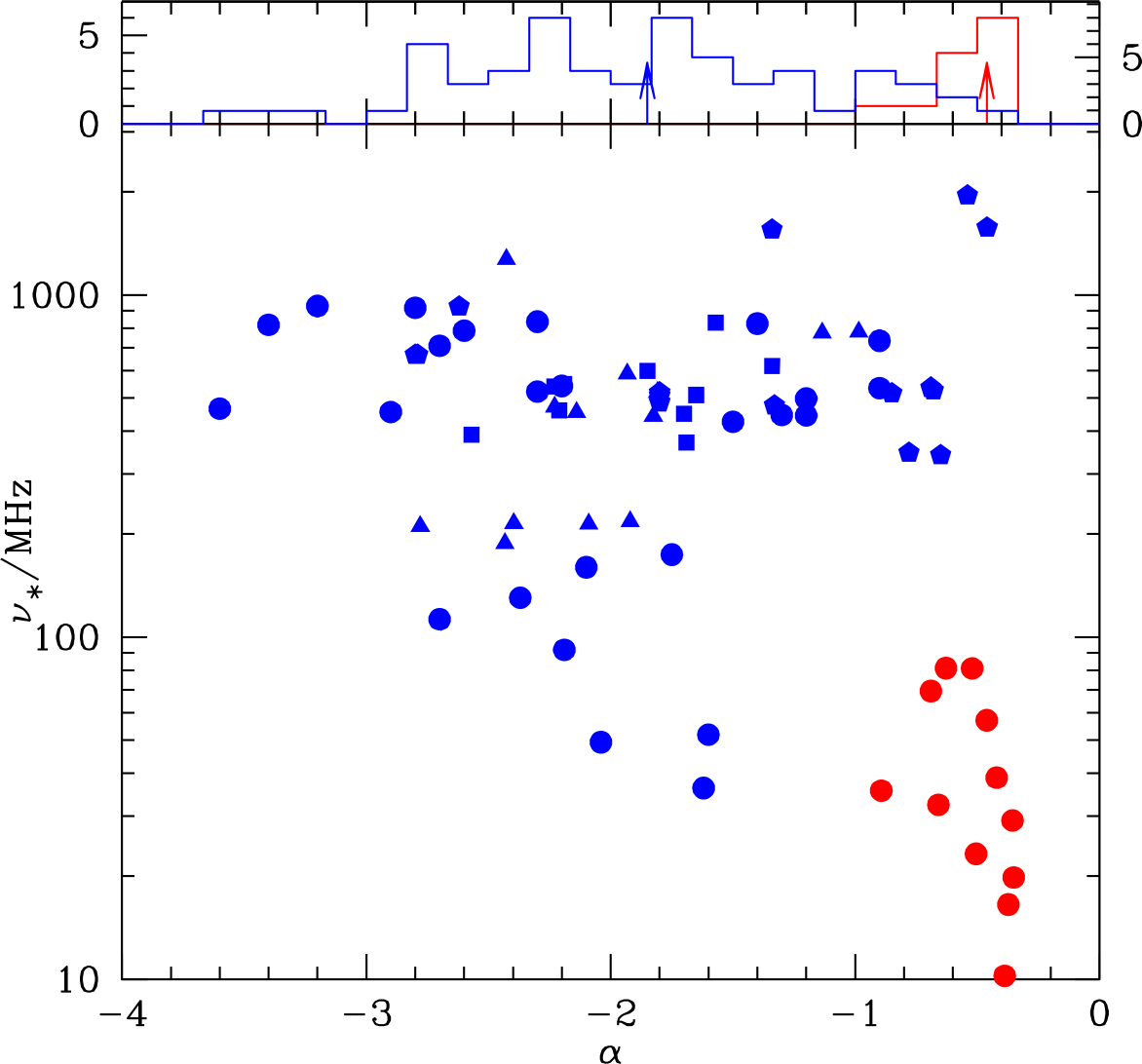}
   \caption{Characteristic frequency,  $\nu_{\ast}$,  fitted parameter in the model  with an exponential drop-off of index $-2.1$  
   from Eq.~\ref{equation:sast} vs. the spectral index of the radio emission for the PSRs in our sample, as reported in K17, J18, K21, and S22.   
    Data points for PSRs are shown with blue symbols, as in Fig.~\ref{figure:fig4}, and data 
 for SNR spectra with exponential turnovers from \citet{abadi2024} are shown as red circles. No correlation is evident between the plotted parameters. 
    The histograms in the top panel  show the distribution of $\alpha$ values for the collection of PSRs (blue) and SNRs (red), with  median values marked by arrows.
}
    \label{figure:fig6}
\end{figure}

 \begin{equation}
 \mathcal{L}= 4\pi\,d^{2} \, \frac{S_{\ast}\,\nu_{\ast}}{(\alpha+1)}\; (x_{\mathrm{max}}^{\alpha+1}-x_{\mathrm{min}}^{\alpha+1}),
     \label{equation:lumi}
 \end{equation}

\noindent
Notice that in Eq.~\ref{equation:lumi}, no correction factor is included for the pulsar emission beam solid angle or pulse duty cycle, as these are typically challenging to determine \citep{lorimer2004}. Moreover, it is a common practice to define a pseudo-luminosity at a specified frequency, usually $\nu=1400$~MHz, as 
$\mathcal{L}_{\nu} \sim d^2 S(\nu)$, without explicitly integrating the power law between the minimum and maximum frequencies involved. The relationship between luminosity and spectral index  is shown in Fig.~\ref{figure:lalpha}, with a solid blue curve for PSRs and a solid red curve for SNRs. 
We have adopted  $x_{\mathrm{min}}=0.01$ and $x_{\mathrm{max}}=1000$, following the frequency coverage seen in Fig.~\ref{figure:fig2}. These curves are scaled using the median values for the parameters $S_{\ast}$, $\nu_{\ast}$, and $d$ for PSRs and SNRs (0.024~Jy, 500~MHz, 4.06~kpc)
and (140~Jy, 36~MHz, 7~kpc), respectively. Since independent heliocentric distance measurements, such as those based on parallax, associations with other objects,  or neutral hydrogen absorption measurements (especially for low-latitude pulsars), are not available for all PSRs in our collection, 
we used the distances reported in the Australia Telescope National Facility (ATNF) Pulsar Catalogue \citep{manchester2005},\footnote{Web address \url{http://www.atnf.csiro.au/research/pulsar/psrcat}} computed  from dispersion measure (DM) estimates assuming the NE2001 model for the Galactic distribution of free electrons \citep{cordes2002, cordes2003}. 
As can be seen in the blue and red curve, this analytically  predicted luminosity decreases systematically with the spectral index until it reaches a critical value $\alpha_{0}$, which can be computed numerically by solving:

\begin{figure} 
  \includegraphics[width=0.45\textwidth]{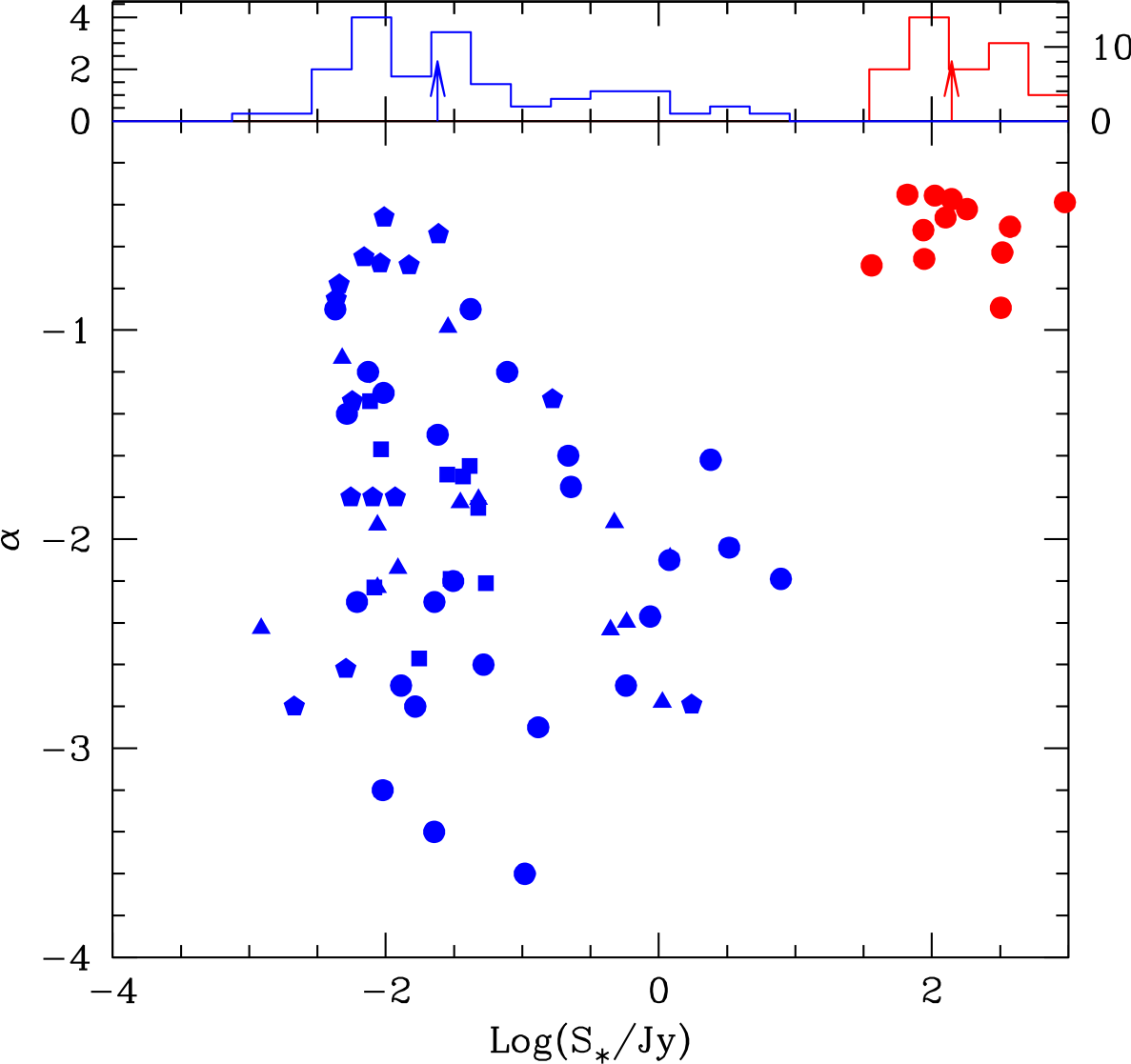}  
\caption{Power-law slope, $\alpha$, vs. the characteristic flux density,  $S_{\ast}$, fitted parameter in  Eq.~\ref{equation:sast} for our collection of 63 PSRs. The spectral index values are those reported in K17, J18, K21, and S22. Blue symbols coded as in Fig.~\ref{figure:fig4}. Red symbols denote SNR data from 
    exponential turnover fits as presented in \citet{abadi2024}. 
    No correlation is evident between the plotted parameters. 
    The histograms in the top panel  illustrate the distribution of $S_{\ast}$ values for PSRs (blue) and SNRs (red), with  median values marked by arrows.}
    \label{figure:fig7}
\end{figure}

\begin{equation}
x_{\mathrm{max}}^{\alpha+1}\, \mathrm{log_{e}}(x_{\mathrm{max}})- x_{\mathrm{min}}^{\alpha+1}\, 
\mathrm{log_{e}}(x_{\mathrm{min}}) = (\alpha+1)^{-1}\, (x_{\mathrm{max}}^{\alpha+1}-x_{\mathrm{min}}^{\alpha+1}).
\end{equation}
\begin{figure}
   \includegraphics[width=0.45\textwidth]{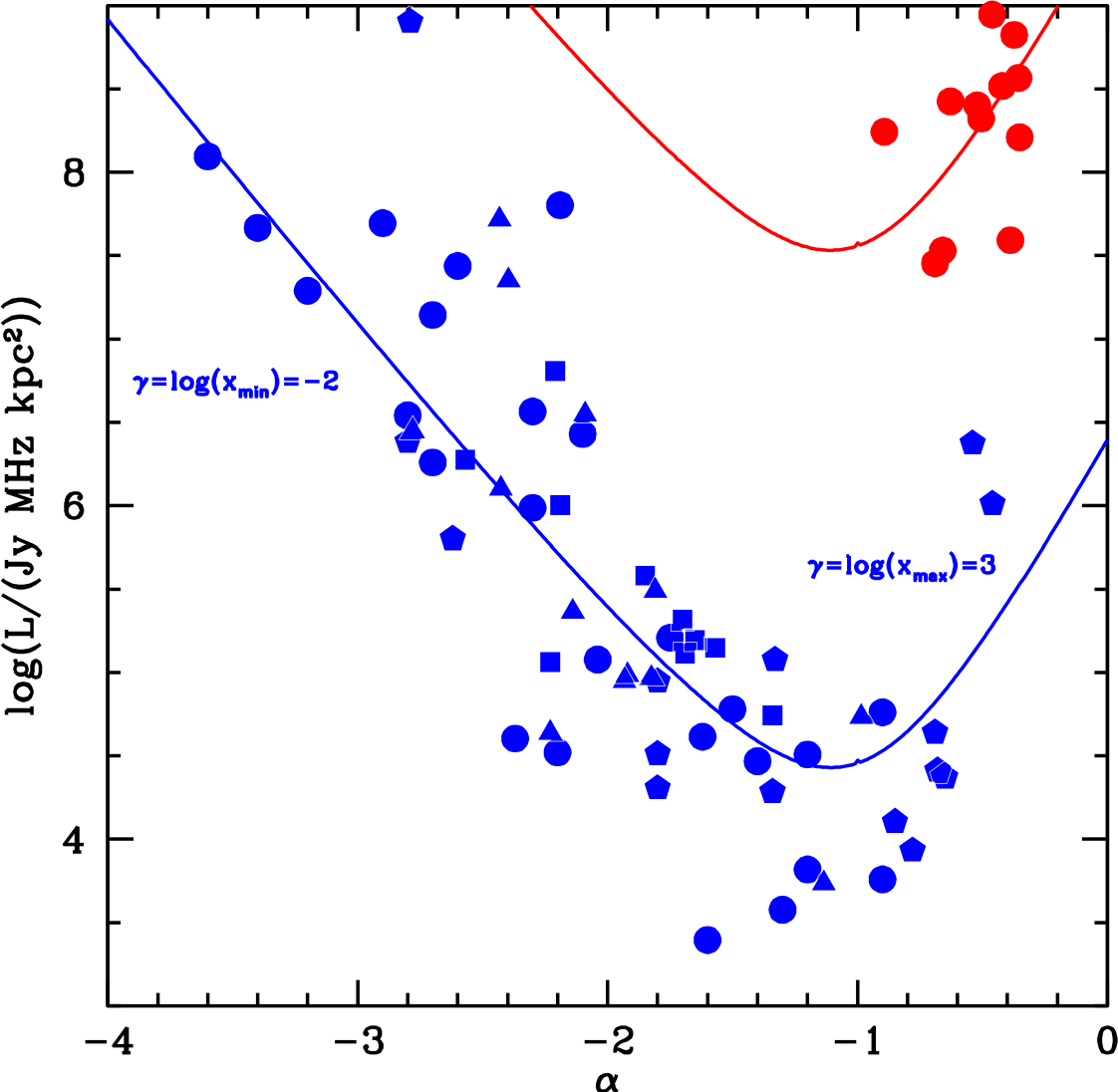}   
    \caption{Integrated luminosity plotted against the power-law spectral index,  $\alpha$,  for our sample of PSRs (blue symbols, as in Fig.~\ref{figure:fig4}) and SNRs (red symbols, from \citealt{abadi2024}). The luminosity is computed by integrating the 
    power-law flux density emission between normalised frequencies $x_{\mathrm{min}}=10^{-2}$ and $x_{\mathrm{max}}=10^{3}$. Notice that changing the frequency range over which  the integration is performed modifies the asymptotic logarithmic slopes, $\gamma$, at the low and high ends of $\alpha$, corresponding to  $\mathrm{log}(x_{\mathrm{min}})$ and $\mathrm{log}(x_{\mathrm{max}})$, respectively. The solid blue and red curves are derived using Eq. (8),  with pre-factor parameters $S_{\ast}, \nu_{\ast}$,  and $d $ given by the corresponding median values for PSRs and SNRs. 
    }
    \label{figure:lalpha}
\end{figure}
For the adopted  $x_{\mathrm{min}}$ and $x_{\mathrm{max}}$ values, we obtain  $\alpha_{0}\sim-1.1$. 
Expanding the integration  interval has a minimal impact (e.g., increasing the range from $x_{\mathrm{min}}=10^{-3}$ to $x_{\mathrm{max}}=10^{3}$ results in  $\alpha_{0}\sim-1$).
The asymptotic logarithmic slopes for spectral indices steeper and flatter than $\alpha_{0}$ are
$\gamma_{<\alpha_{0}}=\mathrm{log}(x_{\mathrm{min}})=-2$ and   $\gamma_{>\alpha_{0}}=\mathrm{log}(x_{\mathrm{max}})=3$, respectively. Using Eq.~\ref{equation:lumi}, we  estimated  the luminosity for each PSR in our sample  and for the collection of SNRs with spectral turnovers.  
The wide range of spectral indices suggests a possible trend where luminosity data points decrease and then increase  with the spectral indices, following the analytical curve. However, the limited number of pulsars with flatter spectral indices than $\alpha_{0}$ precludes a definitive confirmation of this trend.

\section{Physical conditions in the thermal free-free absorbers}
\label{sec:absorbing-medium}
\begin{figure}
\includegraphics[width=0.45\textwidth]{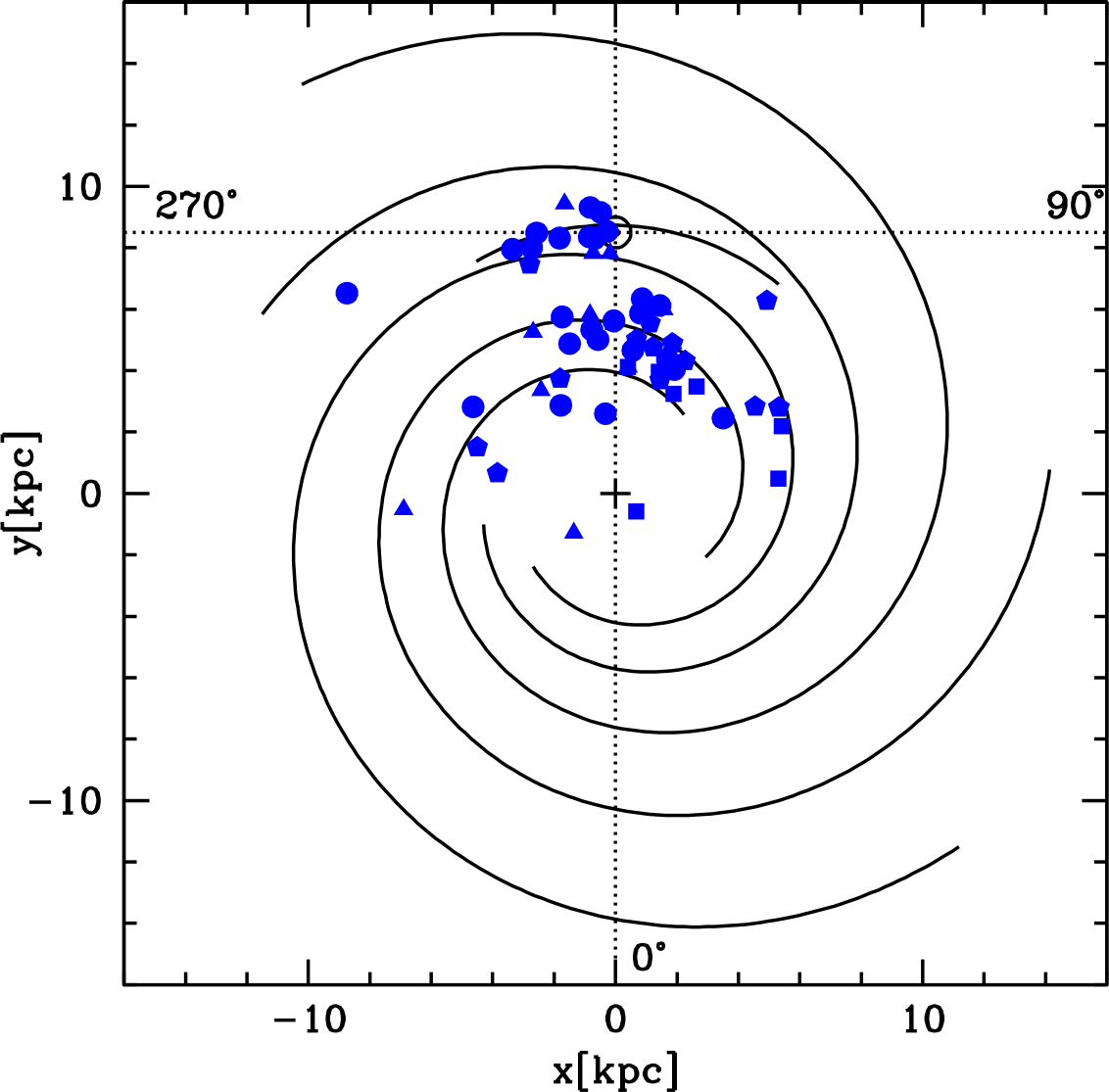}
    \caption{Projected spatial distribution of our sample of 63 PSRs onto the Galactic  plane.
    Distance estimates for the PSRs were taken from the ATNF catalogue \citep{manchester2005} and are based on the NE2001 Galactic electron-density model \citep{cordes2002, cordes2003}. Pentagons correspond to K17 data, circles to J18, squares to K21, and triangles are used to represent S22 data.
    For guidance, the solid black lines show the positions of the spiral arms according to the  \citet{Hou&Han2014} model. The central plus symbol marks the position of the Galactic  centre, the solar symbol shows the Sun's position, and dotted lines outline the Galactic quadrants. 
       }
\label{figure:spiralarmspsr}
\end{figure}

Building on the results presented in the previous sections, we now focus on the physical conditions of the ionised medium responsible for the thermal free-free absorption observed in the PSR radio spectra. As a starting point, we constructed a face-on view of the Galactic plane showing the spatial distribution of the PSRs in our sample (Fig.~\ref{figure:spiralarmspsr}), using  distances inferred from the NE2001 model. 
This representation enables us to infer the likely locations of the absorbers responsible for the observed spectral turnovers, based on the positions of the PSRs, and provides  insight into their distribution relative to large-scale Galactic structures. Although the PSRs in our sample are primarily distributed within the first and fourth Galactic quadrants and extend up to $\sim8$~kpc—thus reaching the vicinity of the Galactic centre—this spatial coverage is sufficient to derive representative  physical properties of the absorbing media.

\begin{figure}
   \includegraphics[width=0.45\textwidth]{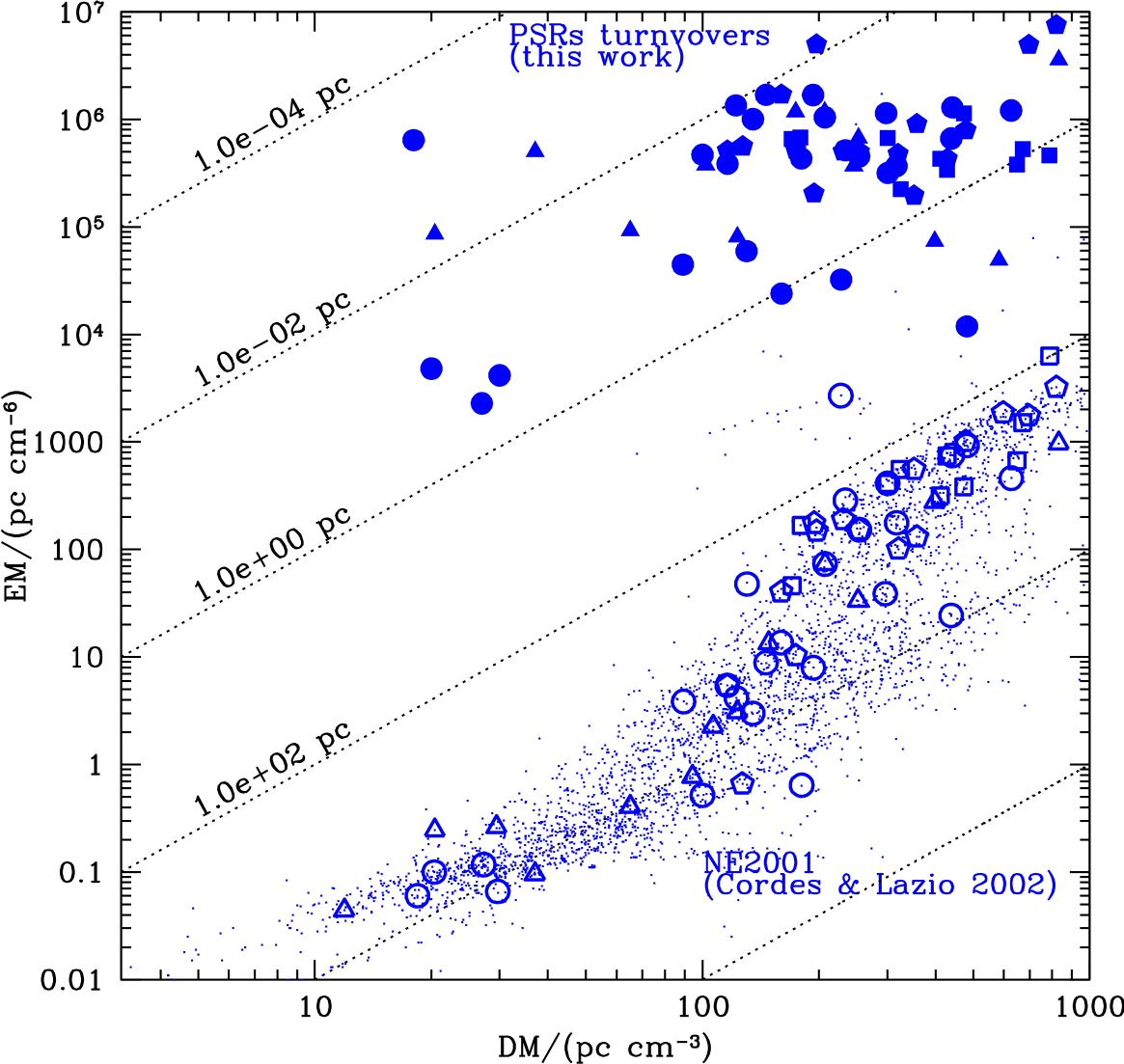}   
    \caption{Emission measures (EM)  plotted against dispersion measures (DM) for the 63 pulsars included in our sample. Filled symbols correspond to EM values derived from radio continuum spectral turnovers (pentagons: K17, circles: J18, squares: K21, and triangles: S22). Open blue symbols represent EM values predicted by the NE2001 model \citep{cordes2002, cordes2003} for our collection of PSRs, while the  smallest blue points indicate values derived  for all PSRs in the ATNF catalogue. 
    The diagonal dotted lines show the relationship for constant size values, as labelled. 
    } 
    \label{figure:emdm}
\end{figure}

To further constrain the properties of the ionised medium, we combined the DMs reported in the ATNF Pulsar Catalogue \citep{manchester2005}  with the EMs derived from our spectral fits, using the relation given in Eq.~\ref{equation:nuast}. For this calculation, we adopted a representative electron temperature of $\mathrm{T_{e}} = 8000$~K, consistent with typical values in 
H\,{\sc ii} regions and other diffuse ionised environments (e.g., \citealp{quireza06,luisi2019}). 
Figure~\ref{figure:emdm} presents the resulting EM--DM diagram, where each filled symbol corresponds to a PSR in our sample. The location of the data points in Fig. ~\ref{figure:emdm} confirms that the observed spectral turnovers correspond to exceptionally high emission measures. 
Given that our sample has a relatively high median turnover frequency ($\nu_{\ast}\approx 500$~MHz, that is $\nu_{\mathrm{to}}\approx 558$~MHz), the resulting EMs are correspondingly large, reaching values up to 
$\sim 10^{5}$~pc~cm$^{-6}$. 
These values are consistent with those reported by \citet{rajwade2016} for a  sample of six  pulsars with GHz spectral turnovers (see their Table~1).

To further highlight the extremely high values inferred from such high-frequency turnovers, we also show the NE2001 model estimates for each PSR in our sample (open symbols in Fig.~\ref{figure:emdm}). 
The NE2001 model assumes a smoothly distributed ionised medium along the line of sight and estimates EM (Eq.~\ref{equation:EM}) based on its internal electron density distribution. In addition, Fig.~\ref{figure:emdm} displays the broader EM vs. DM distribution predicted by NE2001 for the ATNF pulsar population (shown as blue dots). Notably, the PSRs in our sample appear to constitute a representative subsample of this  general  population, showing no  evident bias towards particular values of either EM or DM, as indicated by the overlap between the open symbols and the cloud of blue dots. In contrast to the relatively tight EM–DM correlation expected from a smooth medium, the EM values derived from our spectral fits show no clear trend with DM. This discrepancy suggests that the absorbing material traced by the spectral turnovers is likely dominated by compact and possibly clumpy structures, rather than by a uniformly distributed ionised component. In this sense, our results reinforce the interpretation of thermal absorption as a localised phenomenon shaped by environmental inhomogeneities along the line of sight.  

Combining EM values derived from spectral analysis with DMs obtained from timing observations allows us to disentangle the degeneracy between electron density and absorber size. 
Both, EM and DM can be expressed as $\mathrm{EM}=\int_{0}^{L} n_{\mathrm{e}}^{2} \; dx = n_{\mathrm{e}}^{2}\,L$  and 
$\mathrm{DM}=\int_{0}^{L} n_{\mathrm{e}} \; dx= n_{\mathrm{e}}\,L$, assuming a constant electron density along the path length $L$. 
From these relations, we can estimate the electron density $n_{\mathrm{e}}=\mathrm{EM/DM}$ and the physical size $L=\mathrm{DM}^{2}/\mathrm{EM}$ of the absorbing region. Then, under this simplified model, lines of constant absorber size $L$ in the EM–DM diagram follow a power-law with logarithmic slope 2, with the intercept increasing with $L$, as illustrated by the dotted lines in Fig.~\ref{figure:emdm}. 
Applying these expressions to our sample yields small absorber sizes, most of them typically in the range 0.01 to 1~pc, balanced by high electron densities of a few times $10^{3}$~cm$^{-3}$. 
In contrast, the EM values derived using the NE2001 model correspond to much lower electron densities ($\sim 10^{-1}$~cm$^{-3}$) integrated throughout much larger sizes corresponding to the PSR distance. 
Notably, high-density, compact absorbers have been previously proposed by several authors as the cause of GHz-frequency spectral turnovers in pulsars. For example, \citet{lewandowski2015}, \citet{rajwade2016}, \citet{kijak2017}, \citet{kijak2021} all suggest that such dense, localised structures—such as SNR filaments, bow shocks, or compact H\,{\sc ii} regions—can imprint exponential absorption signatures on the radio spectra of PSRs.

\begin{figure}
   \includegraphics[width=0.45\textwidth]{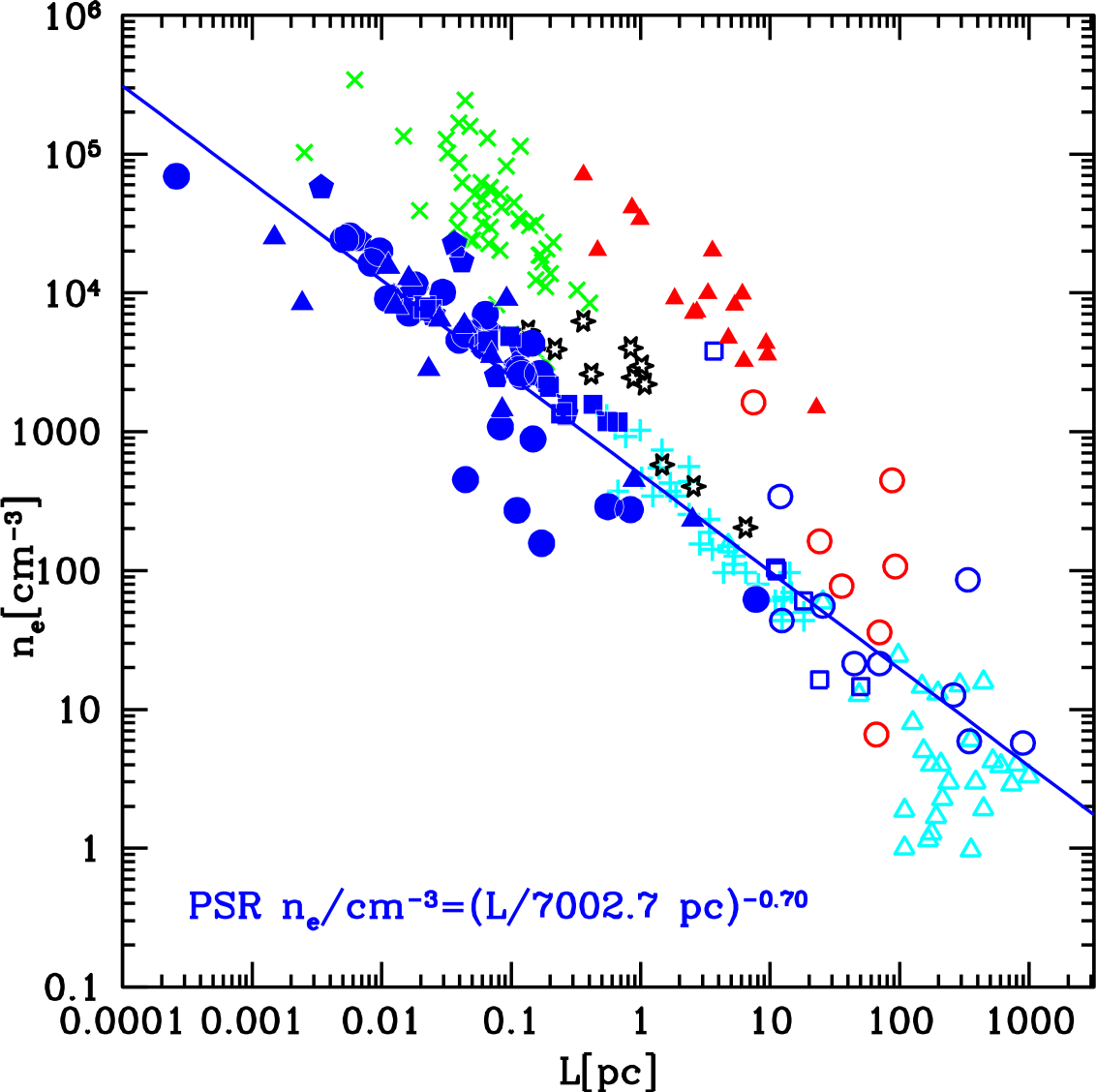}   
    \caption{Relation between the electron number density,  $n_{\mathrm{e}}$, and the size, $L$, of the absorber population  causing turnovers at $\nu_{\ast}< 2$~GHz in our sample of 63 PSR radio continuum spectra,  modelled by a power-law with a turnover due to free-free thermal absorption. Filled blue symbols represent data from PSR spectra in K17 (pentagons), J18 (circles), K21 (squares), and S22 (triangles). 
    The solid line represents the least-squares fit to the PSR data, $n_{\mathrm{e}} \propto L^{-0.70}$, assuming an electron temperature of the ionised gas $\mathrm{T_{e}}= 8000$~K.
   The graphic also reproduces the data points 
   from \citet{hunt2009} (Fig.~2), corresponding to Galactic ultracompact (cyan plus symbols), compact (green crosses), and giant (open cyan triangles) H\,{\sc ii} regions, along with different extragalactic H\,{\sc ii} region samples (radio galaxies, filled red triangles; Small and Large Magellanic Clouds, black open stars; M33, open blue squares; Zw18, filled blue triangles; dwarf galaxies observed with HST, open red and blue circles; spiral and irregular galaxies, open cyan triangles).
   The correlation we found between the size and density of the PSR radio emission absorbers is consistent with that observed ($n_{\mathrm{e}} \propto L^{-1}$) in Galactic and extragalactic H\,{\sc ii} regions. 
    }
    \label{figure:fig10}
\end{figure}

 Further analysis presented in Fig.~\ref{figure:fig10} shows the correlation between $n_{\mathrm{e}}$ and $L$, revealing a simple scaling law in which the electron density decreases as the size of the absorbers increases. The characteristic densities of approximately  $5 \times 10^{3}$~cm$^{-3}$, combined with sizes around $5 \times 10^{-2}$~pc, indicate that the absorbers are both compact and extremely dense. The solid blue line in Fig.~\ref{figure:fig10} represents the best-fit regression solution for our  data points 
(filled blue symbols),  corresponding to 
$n_{\mathrm{e}}=\Big(\frac{L}{7002.7\,\mathrm{pc}}\Big)^{-0.70}$~cm$^{-3}$,  with a highly significant correlation coefficient {$r=-0.86$}. Notably, similar power-law relationships between the sizes and electron densities of ionised regions ($n_{\mathrm{e}} \propto D^{-1}$, where  $D$ is the diameter of the ionised regions; \citealt[][and references therein]{hunt2009})  have been found in studies of Galactic and extragalactic H\,{\sc ii} region populations. In Fig.~\ref{figure:fig10}, we compare the scaling observed in our pulsar sample with the trend reported in H\,{\sc ii} regions, reproducing the data points used by \citet{hunt2009} {in Fig.~2}  of their work. Therefore, our result indicates that the radio continuum radiation from pulsars is absorbed by individual clumps, which follow an electron density-size relationship akin to that of the H\,{\sc ii} regions. 
We notice that both the electron density squared and the absorber size scale with $\mathrm{T}_{\mathrm{e}}^{1.35}$ (see Eq.~\ref{equation:nuast}). This implies that relaxing the assumption of $\mathrm{T_{e}}=8000$~K to a lower (or higher) temperature would result in a corresponding decrease (or increase) in the absorber’s electron density,  while the absorber size would conversely increase (or decrease). Despite these variations in temperature assumptions, the observed correlation remains consistent across different scenarios.

\section{Summary}
\label{sec:conclusions}
We have statistically analysed a sample of 63 PSR radio continuum spectra drawn from the literature, each modelled with a power-law and an exponential turnover attributed to thermal free-free absorption in the intervening ionised gas between the pulsar and the observer. 
Our key findings are summarised as follows:

\begin{enumerate}
    \item We introduced a parameterisation of PSR radio continuum spectra in terms of a characteristic frequency, $\nu_{\ast}$, and its corresponding flux density, $S_{\ast}$, which enables the  homogenisation of different subsamples drawn from previously published studies. This approach eliminates the reliance on an  arbitrarily chosen  reference frequency, allowing for consistent comparison across heterogeneous datasets.

    \item By utilising $\nu_{\ast}$ and $S_{\ast}$ and normalising each spectrum to its intrinsic  power-law emission, we showed that the stacked flux density measurements accurately trace the drop-off caused by thermal absorption, reaching flux density levels that are approximately two orders of magnitude below the flux density at the spectral turnover.

\item Parameterising the PSR spectra revealed a correlation between characteristic flux density and frequency, with a slope  that reflects the median radio spectral index ($\alpha \sim -2.0$) of the PSR sample. We further reinforced this behaviour in the case of SNRs,  which exhibit  significantly flatter  spectra ($\alpha \sim -0.5$).

\item  By combining the EMs  inferred from PSR radio spectral turnovers  with  independently determined DM from pulse time delays,  we estimated  characteristic electron densities and path lengths for the putative  population of absorbers. The resulting values point to a population of clumpy, near-in absorbers, with compact sizes on the order of $L\sim0.1$~pc and relatively higher electron densities of $n_{\mathrm{e}}\sim 10^{3}$~cm$^{-3}$, then as measured towards SNRs.

\item The absorbing structures follow a size-density anticorrelation of the form $n_{\mathrm{e}}\propto L^{-0.70}$, reminiscent of the well-established empirical relation observed in Galactic and extragalactic H\,{\sc ii} regions.
    \end{enumerate}

\begin{acknowledgements}
We thank a very constructive and insightful report from an anonymous referee that improved the previous version of the paper. We also thank Dale Frail and Stella Ocker for a careful reading and comments on a previous version of the paper.
This work has been partially supported by the Consejo de
Investigaciones Cient\'ificas y T\'ecnicas de la Rep\'ublica Argentina
(CONICET, PIP~11220220100337), the Secretar\'ia de Ciencia y T\'ecnica de la
Universidad Nacional de C\'ordoba (SeCyT), Agencia Nacional de 
Promoci\'on Cient\'ifica y Tecnol\'ogica 
(PICT 2019-1600), Argentina. Basic research in radio astronomy at the Naval Research Laboratory is funded by 6.1 Base funding.

\end{acknowledgements}

\bibliographystyle{aa}
\bibliography{Bibliography}

\begin{appendix}
\onecolumn
\section{PSR radio-continuum spectral parameters}

\FloatBarrier 
\label{tab:long}
\begin{longtable}{|r|r|r|r|r|r|r|r|}
\caption{PSR radio-continuum spectral parameters}\\
\hline\hline
\multicolumn{1}{|c|}{PSR} & \multicolumn{1}{|c|}{$l$} & \multicolumn{1}{|c|}{$b$} & \multicolumn{1}{|c|}{$S_{\ast}$} & \multicolumn{1}{|c|}{$\nu_{\ast}$} & \multicolumn{1}{|c|}{$\alpha$} & \multicolumn{1}{|c|}{DM} & \multicolumn{1}{|c|} {d} \\
\multicolumn{1}{|c|}{Name} & \multicolumn{1}{|c|}{[$^\circ$]} & \multicolumn{1}{|c|}{[$^\circ]$} & \multicolumn{1}{|c|}{[mJy]} & \multicolumn{1}{|c|}{[MHz]} & \multicolumn{1}{|c|}{} & \multicolumn{1}{|c|}{[pc~cm$^{-3}$]} & \multicolumn{1}{|c|}{[kpc]} \\

\hline
\endfirsthead
\caption{continued.}\\
\hline
\multicolumn{1}{|c|}{PSR} & \multicolumn{1}{|c|}{$l$} & \multicolumn{1}{|c|}{$b$} & \multicolumn{1}{|c|}{$S_{\ast}$} & \multicolumn{1}{|c|}{$\nu_{\ast}$} & \multicolumn{1}{|c|}{$\alpha$} & \multicolumn{1}{|c|}{DM} & \multicolumn{1}{|c|} {d} \\
\multicolumn{1}{|c|}{Name} & \multicolumn{1}{|c|}{[$^\circ$]} & \multicolumn{1}{|c|}{[$^\circ]$} & \multicolumn{1}{|c|}{[mJy]} & \multicolumn{1}{|c|}{[MHz]} & \multicolumn{1}{|c|}{} & \multicolumn{1}{|c|}{[pc~cm$^{-3}$]} & \multicolumn{1}{|c|}{[kpc]} \\
\hline
\endhead
\hline
\endfoot

J1056$-$6258 & 290.292 & $-2.966$ & 166.60 & 475.6 & $-1.33$ & 320.6 & 2.980 \\
J1550$-$5418 & 327.237 & $-0.132$  & 9.73 & 1576.0 & $-0.46$  & 697.0 &  8.318 \\
J1622$-$4950 & 333.849 & $-0.102$ & 24.36& 1958.0 & $-0.54$ &  820.0 & 8.732 \\
J1644$-$4559 & 339.193 & $-0.195$ &  1741.00 & 669.1 & $-2.79$ & 478.7 & 5.085 \\
J1723$-$3659 & 350.682 & $-0.409$ & 8.02 & 518.7 & $-1.80$ & 254.4 & 3.537 \\
J1809$-$1917 &  11.094 & 0.080 &5.70 & 1553.0 &  $-1.34$ & 197.1 & 3.547 \\
J1825$-$1446 & 16.805 & $-1.001$ & 6.93& 340.3 & $-0.65$ & 352.2 & 5.023 \\
J1826$-$1334 & 18.001 & $-0.691$ &9.11 & 526.5 & $-0.68$ & 231.0 & 3.934 \\
J1830$-$1059 & 20.812 & $-0.478$ & 5.11& 925.5 & $-2.62$ & 159.7 & 3.157 \\
J1835$-$1020 & 21.980 & $-1.302$ & 11.75& 507.8 & $-1.80$ & 115.9  & 2.331 \\
J1841$-$0345 & 28.424 & 0.437 & 4.56&  346.5& $-0.78$ & 194.3 & 4.753 \\
J1852$-$0635 & 27.224 & $-3.341$ & 14.84& 536.1 & $-0.69$ & 173.9 & 4.055 \\
J1901+0510 & 8.737 & 0.025& 5.54& 486.3 & $-1.80$ & 435.0 & 7.320 \\
J1907+0918 &43.024 & 0.730 &2.14 & 669.1 & $-2.80$ & 357.7  & 7.793 \\
J2007+2722 & 65.713 & $-2.695$ & 4.34& 516.8 & $-0.85$ & 127.0 & 5.415 \\
\hline
J0711$-$6830 & 279.531 & $-23.280$ & 31.18&  541.9 & $-2.20$ &  18.4  & 0.864   \\
J0809$-$4753 & 263.301 & $-7.957$ & 864.00&  130.3 & $-2.37$ &  228.3  &  0.266  \\
J0908$-$4913 & 270.265 & $-1.019$ &77.54 & 444.3  &  $-1.20$&  180.4  &  2.568  \\
J0922+0638 & 225.420 & 36.392 &2398.00 & 36.2  &  $-1.62$ &  27.3  & 1.161   \\
J0934$-$5249 &  275.690 & $-0.700$ & 104.30 & 465.3  & $-3.60$ &  100.0  & 1.831 \\
J0942$-$5657 & 279.345 & $-2.990$ & 574.90 & 112.7   & $-2.70$ & 159.7   &  3.401  \\
J0943+1631 & 216.609 & 45.380 &217.50 & 51.8  &   $-1.60$ & 20.3   &  0.815  \\
J1001$-$5507 & 280.226 & 0.084 & 227.50 &  174.2 &  $-1.75$ &  130.3  & 2.777   \\
J1017$-$5621 & 282.732 &0.341 & 227.30& 522.1  & $-2.30$ & 438.7   &  8.963  \\
J1057$-$5226 & 285.984 & 6.649 & 3277.00 & 49.3  & $-2.04$ & 29.7   &  0.714  \\
J1512$-$5759 & 320.772 & $-0.108$ & 52.09 & 786.0  & $-2.60$ & 627.5   &   7.334 \\
J1635$-$5954& 327.747   & $-8.313$ & 12.97 &  710.1 & $-2.70$ & 134.9   &  3.256  \\
J1651$-$4246 & 342.458 & 0.923 & 7832.00& 91.8&  $-2.19$ &  482.0    &  5.902  \\
J1658$-$4958 &  337.602 & $-4.554$ & 9.52&  928.8 & $-3.20$ &  193.4 & 3.916 \\
J1705$-$3950 & 346.341 & 0.718 & 4270.00 & 734.8  & $-0.90$ &  207.3  &     3.265 \\
J1723$-$3659 & 350.682 & $-0.409$ & 7.43& 497.9  &  $-1.20$ & 254.4   &  3.537  \\
J1727$-$2739 & 358.945 & 4.053 & 16.46&   917.5 & $-2.80$ & 146.0   &    2.877 \\
J1751$-$3323 & 356.829  & $-3.382$ & 5.20 & 824.4  & $-1.40$  & 296.7   &  5.909  \\
J1803$-$2137 & 8.395 & 0.146 & 41930.00 & 534.4  & $-0.90$ &  233.9  &  3.878  \\
J1806$-$1154 & 17.142  & 4.417 & 22.67 & 817.6   & $-3.40$ &  122.4  & 2.753 \\
J1832$-$0827 & 23.272    & 0.298  &24.03 & 426.0  &  $-1.50$ & 300.9  &  4.850  \\
J1835$-$1020 & 21.980 & $-1.302$ & 9.64&  445.7 &  $-1.30$ &  115.9  &  2.331  \\
J1836$-$1008 & 22.263    & $-1.415$ & 130.80& 454.8&  $-2.90$ & 316.9   &  4.466  \\
J1843$-$0211 & 30.084  & 0.768 & 6.16 & 835.4&  $-2.30$ &  441.7   &  6.991  \\
J1913$-$0440 & 31.307   & $-7.124$ & 1196.00 &  160.0 & $-2.10$ & 89.4   &  2.785  \\
\hline
J1739$-$3023 & 358.085 & 0.336 & 8.23 & 561.9 &  $-2.23$ & 170.5  & 2.912    \\
J1753$-$2502 & 4.274 & 0.512 & 7.68 & 500.8  &  $-1.34$ & 672.0 & 9.105  \\
J1757$-$2421 & 5.281 & 0.054 & 47.59 & 560.1   & $-1.85$  & 179.5   & 4.398   \\
J1822$-$1400 & 17.252 & $-0.176$ & 17.55&  430.1  & $-2.57$ & 649.3 & 4.754   \\
J1823$-$1115 & 19.767 & 0.946 & 36.75 & 403.9   &  $-1.70$ & 428.6 & 5.583   \\
J1832$-$0827 & 23.272 & 0.298 & 29.90& 563.7  & $-2.19$ & 300.9   & 4.850   \\
J1833$-$0827 & 23.386 & 0.063 & 41.17& 453.5  &  $-1.65$  & 410.9 & 4.665   \\
J1841$-$0425 & 27.818 & 0.279 & 28.18 &  329.2 & $-1.69$  & 325.5 & 5.667   \\
J1852+0031 & 33.523 & 0.017 & 54.15& 470.2  &  $-2.21$ & 787.0 & 9.608   \\
J1906+0641 & 40.604 & $-0.304$ & 9.22 & 725.7  & $-1.57$  & 472.8 & 8.317   \\
\hline
J0614$-$3329 & 240.501 & $-21.827$ &4.99 &  470.4 & $-2.23$ &  37.1  &  1.900  \\
J1430$-$6623 & 312.653 & $-5.404$ & 66.89& 217.8  & $-1.92$ & 65.1   &    0.993 \\
J1513$-$5908 & 320.321  & $-1.162$ & 8.26& 587.8  & $-1.93$ &  252.5  &  4.210  \\
J1524$-$5706 & 322.568 & $-0.186$ &7.49 &  1274.0  & $-2.43$ &  832.0  &    11.351 \\
J1633$-$5015 & 334.698 & $-1.569$  &48.71 & 214.8  & $-2.40$  &  398.4  &  5.680  \\
J1705$-$3950 & 346.341 & 0.718  &6.38 &  775.2 & $-1.13$ &  207.3  &   3.265 \\
J1717$-$3425 & 352.120 &  2.025  & 25.95 &  188.2 &  $-2.43$ &  583.5    &  9.874  \\
J1731$-$4744 & 342.565 & $-7.669$ &138.90 & 214.2  & $-2.09$ & 123.1   &   2.790 \\
J1751$-$4657 & 345.002 & $-10.178$ & 56.94& 210.8  & $-2.78$ &  20.4  &    0.698 \\
J1757$-$2421 & 5.281 & 0.054  & 36.44&   519.4  & $-1.81$ &  179.5  &   4.398 \\
J1831$-$0823 & 23.211 & 0.548  & 6.68&  454.5  & $-2.14$  &  245.9  &    4.450 \\
J1843$-$0000 & 32.013 & 1.769  & 19.86& 441.8  & $-1.83$ & 101.5   &   2.955 \\
J1852$-$0635 & 27.224 &$-3.341$ & 36.89 & 780.2 & $-0.98$ & 173.9 &    4.055 \\

\end{longtable}

\tablefoot{Each group of rows, separated by  horizontal lines, corresponds to data from K17, J18, K21, and S22,  respectively. Columns 1-3 provide the pulsar name (in J2000 coordinates) and  Galactic coordinates. Columns 4 and 5 list the computed values  for $\nu_{\ast}$  and $S_{\ast}$, respectively (see Eq.~\ref{equation:sast}), while Col.~6 reports the radio spectral index from the literature. The last two columns present the DM estimates from the ATNF catalogue and the pulsar distances derived from the NE2001 model.}

\FloatBarrier 
\clearpage

\end{appendix}

\end{document}